\documentclass[nohyperref]{article}

\usepackage{microtype}
\usepackage{graphicx}
\usepackage{subfigure}
\usepackage{xspace}
\usepackage[flushleft]{threeparttable}
\usepackage{amsfonts}
\usepackage{amsmath}
\usepackage{amsthm}
\usepackage{mathtools}
\usepackage{bbm}
\usepackage{enumitem}

\theoremstyle{plain}
\newtheorem{theorem}{Theorem}[section]
\newtheorem{proposition}[theorem]{Proposition}

\theoremstyle{definition}

\newtheorem{assumption}[theorem]{Assumption}
\theoremstyle{remark}
\newtheorem{remark}[theorem]{Remark}

\usepackage[framemethod=TikZ]{mdframed}
\usepackage{amssymb}
\usepackage{multirow, booktabs}
\usepackage{hyperref}

\DeclareMathOperator*{\argmin}{arg\,min}
\newcommand{\norm}[1]{\left\lVert#1\right\rVert}
\newcommand{\abs}[1]{\left\lvert#1\right\rvert}

\newcommand{\bx}{\boldsymbol{x}}
\newcommand{\bs}{\boldsymbol{s}}
\newcommand{\mt}{\mathcal{T}}
\newcommand{\ms}{\mathcal{S}}
\newcommand{\ml}{\mathcal{L}}
\newcommand{\ma}{\mathcal{A}}
\newcommand{\btheta}{\boldsymbol{\theta}}
\newcommand{\bvartheta}{\boldsymbol{\vartheta}}

\newcommand{\eg}{\hbox{{e.g.}}\xspace}
\newcommand{\ie}{\hbox{{i.e.}}\xspace}

\newcommand{\etc}{\hbox{{etc.}}\xspace}

\usepackage{hyperref}

\usepackage[accepted]{icml2022}

\icmltitlerunning{Privacy for Free: How does Dataset Condensation Help Privacy?}

\begin{document}

\twocolumn[
\icmltitle{Privacy for Free: How does Dataset Condensation Help Privacy?}



\icmlsetsymbol{equal}{*}

\begin{icmlauthorlist}
\icmlauthor{Tian Dong}{to,equal}
\icmlauthor{Bo Zhao}{goo}
\icmlauthor{Lingjuan Lyu}{ed}
\end{icmlauthorlist}

\icmlaffiliation{to}{Department of Computer Science and Engineering, Shanghai Jiao Tong University}
\icmlaffiliation{goo}{School of Informatics, The University of Edinburgh}
\icmlaffiliation{ed}{Sony AI}
\icmlcorrespondingauthor{Lingjuan Lyu}
{Lingjuan.Lv@sony.com}

\icmlkeywords{Machine Learning, ICML}

\vskip 0.3in
]



\printAffiliationsAndNotice{\icmlEqualContribution} 

\begin{abstract}

To prevent unintentional data leakage, research community has resorted to data generators that can produce differentially private data for model training.
However, for the sake of the data privacy, existing solutions suffer from either expensive training cost or poor generalization performance.
Therefore, we raise the question whether training efficiency and privacy can be achieved simultaneously. In this work, we for the first time identify that dataset condensation (DC) which is originally designed for improving training efficiency 
is also a better solution to replace the traditional data generators for private data generation, thus providing privacy for free.
To demonstrate the privacy benefit of DC, we build a connection between DC and differential privacy, and theoretically prove on linear feature extractors (and then extended to non-linear feature extractors) that the existence of one 
sample has limited impact ($O(m/n)$) on the parameter distribution of networks trained on $m$ samples synthesized from $n (n \gg m)$ raw 
samples by DC.
We also empirically validate the visual privacy and membership privacy of DC-synthesized data by launching both the loss-based and the state-of-the-art likelihood-based membership inference attacks.
We envision this work as a milestone for data-efficient and privacy-preserving machine learning. 
\end{abstract}

\section{Introduction}

Machine learning models are notoriously known to suffer from a wide range of privacy attacks~\cite{lyu2020threats}, such as model inversion attack~\cite{fredrikson2015model}, membership inference attack (MIA) \cite{shokri2017membership}, property inference attack~\cite{melis2019exploiting}, \etc The numerous concerns on data privacy make it impractical for data curators to directly distribute their private data for purpose of interest.
Previously, generative models, \eg, generative adversarial networks (GANs) \cite{goodfellow2014generative}, was supposed to be an alternative of data sharing.
Unfortunately, the aforementioned privacy risks exist not only in training with raw data but also in training with synthetic data produced by generative models \cite{DBLP:conf/ccs/ChenYZF20}.
For example, it is easy to match the fake facial images synthesized by GANs with the real training samples from the same identity \cite{webster2021person}.
To counter this issue, existing efforts \cite{xie2018differentially, wang2021dpgen, cao2021don, harder2021dp} applied differential privacy (DP) \cite{dwork2006calibrating}
to develop differentially private data generators (called DP-generators), because DP is the \textit{de facto} privacy standard which provides theoretical guarantees of privacy leakage. 
Data produced by DP-generators can then be applied to various downstream tasks, \eg, data analysis, visualization, training privacy-preserving classifier, \etc


However, due to the noise introduced by DP, the data produced by DP-generators are of low quality, which impedes the utility as training data, \ie, accuracy of the models trained on these data.
Thus, more data generated by DP-generators are needed to obtain good generalization performance, which inevitably decreases the training efficiency. 

\begin{figure}[t]
    \centering
    \includegraphics[width=\linewidth]{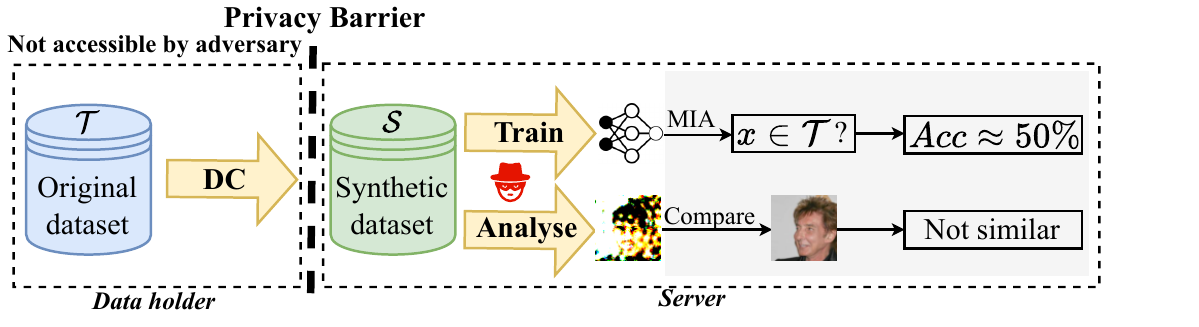}
    \vspace*{-5mm}
    \caption{DC-synthesized data can be used for privacy-preserving model training and cannot be recovered through MIA and visual comparison analysis.}
    \label{fig:overview}
\end{figure}

Recently, the research of dataset condensation (DC) \cite{dataset_distillation, sucholutsky2019soft, such2020generative, bohdal2020flexible, DC_gradient_matching, DSA, DC_distribution_matching, KIP_iclr21, KIP_neurips21, jin2022graph, cazenavette2022distillation, wang2022cafe} emerges, which aims to condense a large training set into a small synthetic set that is comparable to the original one in terms of training deep neural networks (DNNs).
Different from traditional generative models that are trained to generate real-looking samples with high fidelity, these DC methods generate informative training samples for data-efficient learning.
In this work, we for the first time investigate the feasibility of protecting data privacy using DC techniques.
We find that DC can not only accelerate model training but also 
offer privacy for free.
Figure~\ref{fig:overview} illustrates how DC methods
can be applied to protect membership privacy and  visual privacy. 
Specifically, we first analyse the relationship between DC-synthesized data and original ones (Proposition \ref{proposition:DM_linear} and \ref{proposition:minimizer}), and theoretically prove on linear DC extractors that the change caused by removing or adding one element in $n$ raw samples to the parameter distribution of models trained on $m (m\ll n)$ DC-synthesized samples (\ie, privacy loss) is bounded by $O(m/n)$ (Proposition \ref{proposition:KL_div}), which satisfies that one element does not greatly change the model parameter distribution (the concept of DP). 
The conclusions are further analytically and empirically generalized to non-linear feature extractors.
Then, we empirically validate that models trained on DC-synthesized data are robust to both vanilla loss-based MIA and the state-of-the-art likelihood-based MIA \cite{carlini2021membership}.
Finally, we study the visual privacy of DC-synthesized data in case of adversary's direct matching attack.
All the results show that DC-synthesized data are not perceptually similar to the original data as our Proposition \ref{proposition:minimizer} indicates, and cannot be reversed to the original data through similarity metrics (\eg, LPIPS).

Through empirical evaluations on image datasets, we validate that DC-synthesized data can preserve both \textit{data efficiency} and \textit{membership privacy} when being used for model training.
For example, on FashionMNIST, DC-synthesized data enable models to achieve a test accuracy of at least $33.4\%$ higher than that achieved by DP-generators under the same empirical privacy budget.
Meanwhile, to achieve a test accuracy of the same level, DC only needs to synthesize at most $50\%$ data of the size required by GAN-based methods, which speeds up the training by at least $2$ times.

In summary, our contributions are three-fold:
\begin{itemize}
    \item To the best of our knowledge, we are the first to introduce the emerging dataset condensation techniques into privacy community and provide systematical audit on state-of-the-art DC methods.
    \item We build the connection between dataset condensation and differential privacy, and contribute theoretical analysis with both linear and non-linear feature extractors.
    \item Extensive experiments on image datasets empirically validate that DC methods reduce the adversary advantage of membership privacy to zero, and DC-synthesized data are perceptually irreversible to original data in terms of similarity metrics of $L_2$ and LPIPS.
\end{itemize}

\section{Background and Related Work}
In this section, we briefly present dataset condensation and the membership privacy issues in machine learning models.

\subsection{Dataset Condensation}
\label{subsec:DC}
Orthogonal to model knowledge distillation \cite{hinton2015distilling}, \citeauthor{dataset_distillation}
firstly proposed dataset distillation (DD) which aims to distill knowledge from a large training set into a small synthetic set.
The synthetic set can be used to efficiently train deep neural networks with a moderate decrease of testing accuracy.
Recent works significantly advanced this research area by proposing Dataset Condensation (DC) with gradient matching \cite{DC_gradient_matching, DSA}, Distribution Matching (DM) \cite{DC_distribution_matching} and introducing Kernel Inducing Points (KIP) \cite{KIP_iclr21, KIP_neurips21}.
For example, the synthetic sets (50 images per class) generated by DM can be used to train a 3-layer convolutional neural networks from scratch and obtain over $60\%$ testing accuracies on CIFAR10 \cite{krizhevsky2009learning} and over $98\%$ testing accuracies on MNIST \cite{lecun1998gradient}. 
In this work, we mainly focus on synthetic sets generated by DSA \cite{DSA}, DM \cite{DC_distribution_matching} and KIP \cite{KIP_iclr21}, because 1) DSA and DM are improved DC and KIP is improved DD, and 2) the performance of DD and DC are significantly lower than DSA, DM and KIP.


We formulate dataset condensation problem using the symbols presented in \cite{DC_distribution_matching}.
Given a large-scale dataset (target dataset) $\mt=\{(\bx_i, y_i)\}$ which consists of $\abs{\mt}$ samples from $C$ classes, the objective of dataset condensation (or distillation) is to learn a synthetic set $\ms=\{(\bs_i, y_i)\}$ with $\abs{\ms}$ synthetic samples so that the deep neural networks can be trained on $\ms$ and achieve comparable testing performance to those trained on $\mt$:
\begin{equation}
    \mathbb{E}_{\bx \sim P_{\mathcal{D}}}[\ml(\phi_{\btheta^\mt}(\bx),y)]\simeq\mathbb{E}_{\bx \sim P_{\mathcal{D}}}[\ml(\phi_{\btheta^\ms}(\bx),y)], 
\end{equation}
where $P_{\mathcal{D}}$ is the real data distribution, $\phi_{\btheta^\mt}(\cdot)$ and $\phi_{\btheta^\ms}(\cdot)$ are models trained on $\mt$ and $\ms$ respectively. $\ml(\cdot, \cdot)$ is the loss function, \eg cross-entropy loss.

To achieve this goal, \citeauthor{dataset_distillation} proposed a meta-learning based method which parameterizes the model updated on synthetic set as $\btheta^\ms (\ms)$ and then learns the synthetic data by minimizing the validation loss on original training data $\mt$:
\begin{equation}
\argmin_{\mathcal{S}}\ml^{\mt}(\btheta^\ms (\ms)),
\end{equation} 
where $\btheta^\ms(\ms) = \argmin_{\btheta}\ml^\ms(\btheta)$. The meta-learning algorithm has to recurrently unroll the computation graph $\btheta^\ms$ with respect to $\ms$, which is expensive and unscalable. \cite{KIP_iclr21} proposed Kernel Inducing Points (KIP) which leverages the neural tangent kernel (NTK) \cite{neural_tangent_kernel} to replace the expensive network parameter updating. With NTK, $\btheta^\ms$ has a closed-form solution. Thus, KIP learns synthetic data by minimizing the kernel ridge regression loss:
\begin{equation}
   \argmin_{X_s} \frac{1}{2} \norm{y_t-K_{X_t X_s}(K_{X_s X_s}+\lambda I)^{-1}y_s}^2,
\end{equation}
where $X_s$ and $X_t$ are the synthetic and real images from $\ms$ and $\mt$, $y_s$ and $y_t$ are corresponding labels. $K_{UV}$ represents the NTK matrix $(K(u,v))_{(u, v)\in U, V}$ for two sets $U$ and $V$.
For a neural network $\phi_{\boldsymbol\theta}$, the definition of $K(u,v)$ on elements $u$ and $v$ is $
K(u,v) = \nabla_{\boldsymbol\theta} \phi_{\boldsymbol\theta}(u) \cdot \nabla_{\boldsymbol\theta} \phi_{\boldsymbol\theta}(v)$.

\citeauthor{DC_gradient_matching} proposed a novel DC framework to condense the real dataset into a small synthetic set by matching the gradients when inputting real and synthetic batches into the same model, which can be expressed as follows:
\begin{equation}
\argmin_{\ms} \mathrm{E}_{\btheta_0\sim P_{\btheta_0}}[\sum_{t=0}^{T-1} D(\nabla_{\btheta}\ml^\ms(\btheta_{t}),\nabla_{\btheta}\ml^\mt(\btheta_{t}))],
\label{eq:DC}
\end{equation}
where model $\btheta_t$ is updated by minimizing the loss $\ml^\ms(\btheta_{t})$ alternatively, $D$ computes distance between gradients. \cite{DSA} enabled the learned synthetic images 
to be effectively used to train neural networks with data augmentation by introducing the differentiable Siamese augmentation (DSA) $\ma_{\omega}(\cdot)$ and improved the matching loss in \eqref{eq:DC} as follows:
\begin{equation}
D(\nabla_{\btheta}\ml(\btheta_{t}, \ma_{\omega}(\ms)),\nabla_{\btheta}\ml(\btheta_{t}, \ma_{\omega}(\mt))).
\label{eq:DSA}
\end{equation}
Although \cite{DC_gradient_matching} successfully avoided unrolling the recurrent computation graph in \cite{dataset_distillation}, it still needs to compute the expensive bi-level optimization and second-order derivative. To further simplify the learning of synthetic data, \cite{DC_distribution_matching} proposed a simple yet effective dataset condensation method with distribution matching (DM). Specifically, the learned synthetic data $\ms$ should have data distribution close to that of real data $\mt$ in randomly sampled embedding spaces: 
\begin{equation}
\label{eq:loss_DM}
\begin{split}
 \min_{\ms}\mathbb{E}_{\bvartheta \sim P_{\bvartheta}, \omega \sim \Omega}\|\frac{1}{|\mt|}\sum_{i=1}^{|\mt|}\psi_{\bvartheta}(\ma(\bx_i, \omega)) \\ - \frac{1}{|\ms|}\sum_{j=1}^{|\ms|}\psi_{\bvartheta}(\ma(\bs_j, \omega))\|^2,
 \end{split}
\end{equation}
where $\psi_{\bvartheta}$ represents randomly sampled embedding functions (namely feature extractors), \eg randomly initialized neural networks and $\ma(\cdot, \omega)$ is the differentiable Siamese augmentation.
Experimental results show that this simple objective can lead to effective synthetic data that are comparable even better than those generated by existing methods.

\subsection{Membership Privacy}
For privacy analysis, we mainly focus on membership privacy as it directly relates to personal privacy.
For example, inferring that an individual's facial image was in a shop's training dataset reveals the individual had visited the shop.
\citeauthor{shokri2017membership} have shown that DNNs' output can leak the membership privacy of the input (\ie, whether the input belongs to the training dataset) under \textit{membership inference attack} (MIA).
In general, MIA only needs black-box access to model parameters \cite{sablayrolles2019white} and can be successful with logit
\cite{yu2021does} 
or hard label prediction \cite{DBLP:conf/ccs/LiZ21, pmlr-v139-choquette-choo21a}.

\noindent\textbf{Loss-based MIA.} The loss-based MIA infers membership by the predicted loss: if the loss is lower than a threshold $\tau$, then the input is a member of the training data.
Formally, the membership $M(\mathbf{x})$ of an input $\mathbf{x}$ can be expressed as:
\begin{equation}
    M(\mathbf{x}) = \mathbbm{1}(l(\mathbf{x})\leq \tau),
\end{equation}
where $M(\mathbf{x})=1$ means $\mathbf{x}$ is a training member, $\mathbbm{1}(A)=1$ if event $A$ is true.
The threshold $\tau$ can be either chosen by locally trained shadow models \cite{shokri2017membership} or via optimal bayesian strategy \cite{sablayrolles2019white}.

\noindent\textbf{Likelihood-based MIA.} Recent works \cite{carlini2021membership,rezaei2021difficulty} pointed out that the evaluation of MIA should include False Postive Rate (FPR) instead of averaged metrics (\eg, attack accuracy, Area Under Curve (AUC) score of Receiver Operating Characteristic (ROC) curve), because MIA is a real threat only if the FPR is low (\ie, few data are inferred as member).
Moreover, \citeauthor{carlini2021membership} also discovered that loss-based MIAs are hardly effective under constraint of low FPR (\eg, FPR $<0.1\%$).
Hence, they devised a more advanced MIA, \ie Likelihood Ratio Attack (LiRA) based on the model output difference caused by membership of an input.
We consider the \textit{online LiRA attack} because of its high attack performance.
Particularly, the adversary first prepares shadow models ahead of the attack by sampling $N$ sub-datasets and training shadow models on each of the sampled dataset.
Hence, for each data sample, there are $\frac{N}{2}$ shadow models that are trained on it (called IN models) and the rest $\frac{N}{2}$ that are not trained on it (called OUT models).
The adversary then measures the means $\boldsymbol\mu_{in}, \boldsymbol\mu_{out}$ and the variances $\boldsymbol\sigma^2_{in}, \boldsymbol\sigma^2_{out}$ of model confidence for IN and OUT models, respectively.
Here, the confidence of model $f$ for $(\mathbf{x}, y)$ is $\phi(f(\mathbf{x})_y)=\phi(\exp(-l(f(\mathbf{x}), y)))$, where $l$ is the cross-entropy loss and $\phi(p)=\log(\frac{p}{1-p})$.
To attack, the adversary queries the victim model $f$ with a target example $(\mathbf{x},y)$ to estimate the likelihood $\Lambda$ defined as:
\begin{equation}
    \Lambda = \frac{p(\text{conf}_{obs}|\mathcal{N}(\boldsymbol\mu_{in}, \boldsymbol\sigma^2_{in}))}{p(\text{conf}_{obs}|\mathcal{N}(\boldsymbol\mu_{out}, \boldsymbol\sigma^2_{out}))},
\end{equation}
where $\text{conf}_{obs}=\phi(f(\mathbf{x})_y)$ is the confidence of victim model $f$ on target example $(\mathbf{x},y)$.
The adversary infers membership by thresholding the likelihood $\Lambda$ with threshold $\tau$ determined in advance.
\section{Problem Statement}
\label{sec:problem}

In practice, companies may utilize personal data for model training in order to provide better services.
For example, 
data holders (\eg, smart retail stores, smart city facilities) may capture clients' 
data and send to 
cloud servers for model training.
However, models trained on the raw data (\ie, $\mathcal{T}$) can be attacked by MIA.
In addition, transmitting raw data to servers 
suffers from potential data leakage (\eg, to honest-but-curious operators).
Therefore, a better protocol is to first learn knowledge from data by, for instance, generating synthetic dataset $\mathcal{S}$ from the raw data (\ie, $\mathcal{T}$), and then send $\mathcal{S}$ to the server 
for model training for downstream applications.
Formally, we define the threat model as follows:

\noindent\textbf{Adversary Goal.}
The adversary aims to examine the membership 
information of the target dataset $\mathcal{T}$.
Specifically, for a sample of interest $\mathbf{x}$, the adversary infers whether $\mathbf{x}\in\mathcal{T}$.

\noindent\textbf{Adversary Knowledge.}
We assume a strong adversary (\eg, honest-but-curious server), who although has no access to $\mathcal{T}$ but has the \textit{white-box} access to both the synthetic dataset $\mathcal{S}$ synthesized from the target dataset $\mathcal{T}$ and the model $f_\mathcal{S}$ trained on the synthetic dataset.
The adversary also knows the data distribution of $\mathcal{T}$.

\noindent\textbf{Adversary Capacity.}
The adversary has unlimited computational power to generate shadow synthetic datasets on data of same distribution of $\mathcal{T}$ and train shadow models on them.

Note that white-box access to the model parameters does not help MIA \cite{sablayrolles2019white}, so we omit other advantages brought by the white-box access to  $f_\mathcal{S}$.

\section{Theoretical Analysis}
\label{sec:theory}
In this section, we theoretically analyse the relationship between the target dataset $\mathcal{T}$ and the synthetic dataset $\mathcal{S}$ of DM (improved DC)~\cite{DC_distribution_matching}, and the privacy guarantees of $\mathcal{T}$ that are thereby provided.
The reason of choosing DM is because of its high condensation efficiency and utility for model training.
We also verify the 
difference 
between DM and other DC methods (see Appendix \ref{subsec:visual_distribution}) in terms of the synthetic data distribution, indicating our theoretical results can be generalized to other methods to some extent.
Theoretical analysis of other DC methods (\eg, DSA) is left as the future work.

\textbf{Overview.}
We briefly present an overview of the analysis that consists of three parts.
First, we clarify the assumptions and notations in Section \ref{subsec:assumption}.
Then, in Section \ref{subsec:data}, we analyse the connection between synthetic and original datasets for different DC initializations.
Finally, in Section \ref{subsec:model}, with conclusion of Section \ref{subsec:data}, we study the privacy loss of models trained on DC-synthesized data in a DP manner: how does removing one sample in the original dataset impact models trained on synthetic dataset.
Because of the randomness in model training, we base on the model parameter distribution assumption from \cite{sablayrolles2019white} and compute the order of magnitude of the impact, which establishes the connection between DC and DP.

\subsection{Assumptions \& Notations}
\label{subsec:assumption}
The DM loss \eqref{eq:loss_DM} can be optimized for each class \cite{DC_distribution_matching}.
To simplify the notations, we consider 
only one class and omit the label vectors in the synthetic dataset $\mathcal{S} =\{\mathbf{s}_1,\cdots,\mathbf{s}_{\abs{\mathcal{S}}}\}$ and the target dataset $\mathcal{T} =\{\mathbf{x}_1,\cdots,\mathbf{x}_{\abs{\mathcal{T}}}\}$.
We consider the following two assumptions of the target dataset and the convergence of DM.

\begin{assumption}
\label{assump:subspace}
The linear span of the target dataset $\text{span}(\mathcal{T})$ satisfies
$d_{\mathcal{T}} =\text{dim}(\text{span}(\mathcal{T})) < d$,
where $d$ is the data dimension, $\mathrm{dim}(V)$ represents the dimension of vector space $V$, $\text{span}(\mathcal{T})$ is the vector subspace generated by all linear combinations of $\mathcal{T}$:
\begin{equation}
    \text{span}(\mathcal{T}) \coloneqq \{\sum\limits_{i=1}^{\abs{T}}w_i \textbf{x}_i | 1 \leq i\leq \abs{\mathcal{T}}, w_i\in \mathbb{R}, \textbf{x}_i\in \mathcal{T} \}.
\end{equation}
\end{assumption}
In practice, $d_{\mathcal{T}}$ can be computed as the rank of matrix form of $\mathcal{T}$.
This assumption generally holds for high dimensional data and can be directly verified for common datasets (\eg, CIFAR-10).
Without loss of generality, we consider an orthogonal basis (under inner product of $\mathbb{R}^d$) $\mathcal{E}=\{\mathbf{e}_1, \cdots, \mathbf{e}_d\}$ among which the first $d_\mathcal{T}$ basis vectors $\mathcal{E}_\mathcal{T}=\{\mathbf{e}_1, \cdots, \mathbf{e}_{d_{\mathcal{T}}}\}$ form the orthogonal basis of vector subspace $\text{span}(\mathcal{T})$.


\begin{assumption}
\label{assump:convergence}
[Convergence of DM]. We assume there exists at least one synthetic dataset $\mathcal{S}^\ast=\{\mathbf{s}_1^\ast, \cdots, \mathbf{s}_{\abs{\mathcal{S}}}^\ast\}$ that minimizes \eqref{eq:loss_DM}.
\end{assumption}

\subsection{Analysis of Synthetic Data}
\label{subsec:data}
We first analyse synthetic data by linear extractors and then discuss the generalization to non-linear case (Remark \ref{remark:non_linear_extractor}).

\begin{proposition}[Minimizer of DM Loss]
\label{proposition:DM_linear}
For a linear extractor $\psi_{\boldsymbol\theta}:\mathbb{R}^d\rightarrow \mathbb{R}^k$ such that 
$k<d$, $\boldsymbol\theta \in \mathbb{R}^{k \times d}$, under Assumption \ref{assump:subspace} and \ref{assump:convergence}, the dataset $\mathcal{S}^\ast$ synthesized by DM from the target dataset $\mathcal{T}$ satisfies: 

1) The barycenters of $\mathcal{S}^\ast$ and $\mathcal{T}$ coincide:
\begin{equation}
    \frac{1}{\abs{\mathcal{T}}}\sum\limits_{i=1}^{\abs{\mathcal{T}}} \mathbf{x}_i - \frac{1}{\abs{\mathcal{S}^\ast}}\sum\limits_{i=1}^{\abs{\mathcal{S}^\ast}} \mathbf{s}_i^\ast=\mathbf{0},
\end{equation}
2) $\forall \mathbf{s}_i^\ast\in\mathcal{S}^\ast,
\mathbf{s}_i^\ast = \mathbf{s}_{i,\mathcal{E}_\mathcal{T}}^\ast + \mathbf{s}_{i,\mathcal{E}_\mathcal{T}^\perp}^\ast$, where $\mathbf{s}_{i,\mathcal{E}_\mathcal{T}}^\ast\in \text{span}(\mathcal{T})$, $\mathbf{s}_{i,\mathcal{E}_\mathcal{T}^\perp}^\ast\in \text{span}(\mathcal{T})^\perp$ that verifies
\begin{equation}
    \sum\limits_{i=1}^{\abs{\mathcal{S}}}\mathbf{s}_{i, \mathcal{E}_\mathcal{T}^\perp}^\ast=\mathbf{0}_{ \mathcal{E}_\mathcal{T}^\perp}.
\end{equation}
\end{proposition}

The proof of Proposition \ref{proposition:DM_linear} can be found in Appendix \ref{appendix:proof1}.
Note that the minimizer is $\mathbf{s}^\ast_1 = \frac{1}{\abs{\mathcal{T}}}\sum_{i=1}^{\abs{\mathcal{T}}}\mathbf{x}_i$ when $\abs{\mathcal{S}^\ast}=1$, indicating that the synthetic data 
falls into vector subspace $\text{span}(\mathcal{T})$, confirming the Proposition \ref{proposition:DM_linear}.

The DC initialization of synthetic dataset can be either real data sampled from $\mathcal{T}$ or random noise.
Next, we study the impact of DM initialization and obtain the following results (proof can be found in Appendix \ref{appendix:proof2}).

\begin{proposition}[Connection between $\mathcal{S}^\ast$ and $\mathcal{T}$]
\label{proposition:minimizer}
Based on Proposition \ref{proposition:DM_linear}, the minimizer synthetic dataset $\mathcal{S}^\ast=\{\mathbf{s}_1^\ast,\cdots,\mathbf{s}_{\abs{\mathcal{S}}}^\ast\}$ 
has the following properties for different initialization strategies:

1) Real data initialization. Assume that $\mathcal{S}$ is initialized with first $\abs{\mathcal{S}}$ samples of $\mathcal{T}$, \ie, $\mathbf{s}_i=\mathbf{x}_i$, then we have
\begin{equation}
\label{eq:minimizer_real_data}
\mathbf{s}_i^\ast = \mathbf{x}_i + \frac{1}{\abs{\mathcal{T}}}\sum\limits_{j=1}^{\abs{\mathcal{T}}}\mathbf{x}_j - \frac{1}{\abs{\mathcal{S}}}\sum\limits_{j=1}^{\abs{\mathcal{S}}}\mathbf{s}_j \in \text{span}(\mathcal{T}).
\end{equation}
2) Random initialization.
The synthetic data are initialized with noise of normal distribution, \ie, $\forall \mathbf{s}_i\in \mathcal{S}, \mathbf{s}_i \sim \mathcal{N}(\mathbf{0},\mathbf{I}_d)$, and we assume the empirical mean is zeroed, \ie, $\frac{1}{\abs{\mathcal{S}}}\sum_{i=1}^{\abs{\mathcal{S}}}\mathbf{s}_i=\mathbf{0}$, then we have
\begin{equation}
\label{eq:minimizer_random}
    \mathbf{s}_{i}^\ast = \mathbf{s}_{i, \mathcal{E}_\mathcal{T}}^\ast + \mathbf{s}_{i, \mathcal{E}_\mathcal{T}^\perp}^\ast,
\end{equation}
where 
$
\mathbf{s}_{i, \mathcal{E}_\mathcal{T}}^\ast = \mathbf{s}_{i, \mathcal{E}_\mathcal{T}} + \frac{1}{\abs{\mathcal{T}}}\sum_{j=1}^{\abs{\mathcal{T}}}\mathbf{x}_j \in \text{span}(\mathcal{T}),
$
and
$
\mathbf{s}_{i, \mathcal{E}_\mathcal{T}^\perp}^\ast= \mathbf{s}_{i, \mathcal{E}_\mathcal{T}^\perp} \in (\text{span}(\mathcal{T}))^\perp$.
\end{proposition}

\begin{remark}[Non-linear Extractor]
\label{remark:non_linear_extractor}
Our results can be generalized to the non-linear extractors.
\citeauthor{giryes2016deep} proved that multi-layer random neural networks generate distance-preserving embedding of input data, so \eqref{eq:loss_DM} is minimized if and only if the distance between real and synthetic data is minimized.
Take 2-layer random networks as an example \cite{estrach2014signal}, there exist two constants $0<A\leq B$ such that $\forall (\mathbf{x}, \mathbf{y})\in (\mathbb{R}^d)^2$, $A\norm{\mathbf{x}-\mathbf{y}}_2 \leq \norm{\rho(\btheta\mathbf{x})-\rho(\btheta\mathbf{y})}_2 \leq B \norm{\mathbf{x}-\mathbf{y}}_2$, where $\rho$ is ReLU.
We also analyse the case of 2-layer extractors (activated by ReLU) and found that the (pseudo)-barycenters of $\mathcal{S}^\ast$ and $\mathcal{T}$ still coincide.
Moreover, on convolutional extractors and the 2-layer extractors, we empirically verify Proposition \ref{proposition:DM_linear} in Appendix \ref{appendix:non_linear_extractor} (see Figure \ref{fig:verify_proposition_DM_linear}).
\end{remark}

\begin{remark}[Impact of initialization on privacy]
\label{remark:impact_init_on_privacy}
Note that in case of real data initialization, a higher $\abs{\mathcal{S}}$ results in lower distance between barycenters of initialized $\mathcal{S}$ and $\mathcal{T}$, thus the changes brought to $\mathcal{S}$ become smaller when $\mathcal{S}$ becomes larger.
This explains the phenomenon that DM-generated images are more visually similar to the real images for higher ipc (images per class) \cite{DC_distribution_matching}. 
However, as we demonstrate in our experiments (Section \ref{subsec:membership_privacy_of_fS}), the membership of data used for DC initialization can still be inferred by vanilla loss-based MIA.
One countermeasure is to choose hard-to-infer samples \cite{carlini2021membership}, \ie, samples whose model outputs are not affected by the membership, as initialization data.

On the other hand, data not used for initialization generate little effect (\ie, their weights in synthetic data are $O(\frac{1}{\abs{\mathcal{T}}})$) on synthetic data, just as the case of random initialization, where data in $\mathcal{T}$ also generate little effect on $\mathcal{S}^\ast$.
We demonstrate that the membership of those data cannot be inferred under both loss-based MIA and the state-of-the-art likelihood MIA (see Section \ref{subsec:membership_privacy_of_fS}).
Moreover, projection component of $(\text{span}(\mathcal{T}))^\perp$ can further protect the privacy (\eg, visual privacy in Section \ref{subsec:vision_privacy}).
\end{remark}

\begin{remark}[Comparison between DC and GAN]
The generator of GAN is trained to minimize the distance between the real and the generated data distributions, which is similar to the objective of DC.
However, GAN-generated data share the same constraints (\ie, bounded between $0$ and $1$) as the real data. 
DC-generated data do not need to satisfy these constraints.
This enables the DC-generated data to contain more features and explains the higher accuracy of models trained on DC-generated data \cite{DC_gradient_matching}.
We also empirically compare the accuracy of model trained on GAN-generated data and DC-synthesized data (see Section \ref{subsec:compare_generator})
, and found that DC-synthesized data outperform GAN-generated data for training better models with smaller amount of training data.
\end{remark}

\subsection{Privacy Bound of Models Trained on Synthetic Data}
\label{subsec:model}
To understand how synthetic dataset protects membership privacy of $\mathcal{T}$ when being used for training model $f_{\mathcal{S}}$, we estimate how model parameters change when removing one sample from $\mathcal{T}$ by adopting below assumption.
With a little abuse of notation, we denote the minimizer set $\mathcal{S}^\ast$ by $\mathcal{S}$ when the context is clear.

\begin{assumption}[Distribution of model parameter \cite{sablayrolles2019white}]
\label{assump:model_param_dist}
The distribution of model parameter $\boldsymbol\theta$ given training dataset $\mathcal{T}=\{\mathbf{x}_1,\cdots, \mathbf{x}_{\abs{\mathcal{T}}}\}$ and loss function $l$ is:
\begin{equation}
    \mathbb{P}(\boldsymbol\theta | \mathcal{T}) = \frac{1}{K_{\mathcal{T}}} \exp{(-\sum_{i=1}^{\abs{\mathcal{T}}}l(\boldsymbol\theta, \mathbf{x}_i))},
\end{equation}
where $K_\mathcal{T}$ is the constant normalizing the distribution.
\end{assumption}

Unlike widely known DP mechanisms (\eg, Gaussian mechanism) that transform the deterministic query function into a randomized one, randomness brought by optimization algorithm (\ie, SGD) or hardware defaults leads to different parameters each time of training, which justifies Assumption \ref{assump:model_param_dist} and ensures the ``uncertainty'' in DP.
In addition, we need the following assumption on the datasets $\mathcal{S}$, $\mathcal{T}$ and the loss function $l$ introduced in the Assumption \ref{assump:model_param_dist}.
The assumption is valid for finite datasets and common loss functions (\eg, cross-entropy) and is used to quantify the data bound and loss variation.

\begin{assumption}
\label{assump:lipschitz}
We assume the data of $\mathcal{T}$ and $\mathcal{S}$ are bounded, \ie,
\begin{equation}
    \exists B>0, \forall \mathbf{x} \in \mathcal{T}\cup \mathcal{S}, \norm{\mathbf{x}}_2 \leq B.
\end{equation}
The loss function $l(\boldsymbol\theta, \cdot):\mathbb{R}^d\rightarrow\mathbb{R}^+$ is $L$-Lipschitz according to the $L_2$ norm, \ie,
\begin{equation}
    \forall (\mathbf{x}, \mathbf{y}) \in \mathcal{B}(B)^2, \boldsymbol\theta, \abs{l(\boldsymbol\theta,\mathbf{x}) - l(\boldsymbol\theta,\mathbf{y})} \leq L \norm{\mathbf{x} - \mathbf{y}}_2,
\end{equation}
where $\mathcal{B}(B) = \{\mathbf{x} | \norm{\mathbf{x}} \leq B\}$ is the close ball of space $\mathbb{R}^d$.
\end{assumption}

With all previous assumptions, we have the following result.
\begin{proposition}
\label{proposition:KL_div}
Suppose a target dataset $\mathcal{T}=\{\mathbf{x}_1,\cdots, \mathbf{x}_{\abs{\mathcal{T}}}\}$ and the leave-one-out dataset $\mathcal{T}^\prime=\mathcal{T}\setminus\{\mathbf{x}\}$ such that $\mathbf{x}$ is not used for initialization.
The synthetic datasets are $\mathcal{S}$ and $\mathcal{S}^\prime$ and $\abs{\mathcal{S}}=\abs{\mathcal{S}^\prime} \ll \abs{\mathcal{T}}$.
Denote the model parameter distributions of $\mathcal{S}$ and $\mathcal{S}^\prime$ by $p(\boldsymbol\theta)=\mathbb{P}(\boldsymbol\theta | \mathcal{S})$ and $q(\boldsymbol\theta)=\mathbb{P}(\boldsymbol\theta | \mathcal{S}^\prime)$ respectively.
Then, the membership privacy leakage caused by removing $\mathbf{x}$ is
\begin{equation}
    D_{KL}(p||q) = O(\frac{\abs{\mathcal{S}}}{\abs{\mathcal{T}}}).
\end{equation}
\end{proposition}

Proposition \ref{proposition:KL_div} indicates that the adversary can only obtain limited information (\ie, $O(\frac{\abs{\mathcal{S}}}{\abs{\mathcal{T}}})$) by MIA when the synthetic data is much fewer than the original data ($\abs{\mathcal{S}}\ll \abs{\mathcal{T}}$), which explains why synthetic data $\mathcal{S}$ protects membership privacy of model $f_\mathcal{S}$.
The proof is in Appendix \ref{appendix:proof3}.

\noindent\textbf{Connection to DP.}
Our privacy analysis is based on the impact of model parameter distribution by removing one element from the origin training dataset, which is similar to the definition of DP \cite{dwork2006calibrating}.
Formally, a $\epsilon$-differential privacy mechanism $\mathcal{M}$ satisfies:
\begin{equation}
    \ln\frac{\mathbb{P}(\mathcal{M}(D)\in \mathcal{S}_\mathcal{M})}{\mathbb{P}(\mathcal{M}(D^\prime)\in \mathcal{S}_\mathcal{M})} \leq \epsilon
\end{equation}
for all neighbor dataset pair $(D, D^\prime)$ and all subset $\mathcal{S}_\mathcal{M}$ of the range of $\mathcal{M}$.
Without knowledge of explicit form of model parameter distribution, we can only claim that the privacy budget $\epsilon$ varies at the order of $O(\frac{\abs{\mathcal{S}}}{\abs{\mathcal{T}}})$.

In practice, we use an empirical budget $\hat{\epsilon}$ through MIA \cite{kairouz2015composition} to measure the privacy guarantee against MIA.
Typically, for a MIA that achieves FPR and TPR (True Positive Rate) against a model, the empirical budget is $\hat{\epsilon} \geq \ln (TPR / FPR)$.
In other words, the model behaves $\hat{\epsilon}$-differentially private to an adversary that applies MIA (\ie, threat model in Section \ref{sec:problem}).

Note that the empirical budget $\hat{\epsilon}$ is \textit{not} equivalent to the real budget $\epsilon$ because of different threat models \cite{nasr2021adversary}.
Nonetheless, we consider black-box MIA as the only privacy threat to the model, thus we can regard the DP budget $\epsilon$ as a model privacy metric against MIA.
In this way, we can compare $\hat{\epsilon}$ and $\epsilon$ by the definition of $\hat{\epsilon}$.
In Section \ref{subsec:compare_generator}, we show that models trained on data synthesized by DC achieve $\hat{\epsilon}\approx 2$ against threat from the state-of-the-art MIA (LiRA), and obtain accuracy much higher than differentially private generators \cite{neurips20chen, harder2021dp}, indicating DC is a better option for efficient and privacy-preserving model training.

\section{Evaluation}
In this section, we evaluate the membership privacy of $f_\mathcal{S}$ for real data and random initialization.
Then, we compare DC with previous DP-generators and GAN to demonstrate DC's better trade-off between privacy and utility.
Finally, we investigate the visual privacy of DC-synthesized data.
\subsection{Experimental Setup}
\noindent\textbf{Datasets \& Architectures.}
We use three datasets: FashionMNIST \cite{xiao2017fashion}, CIFAR-10 \cite{krizhevsky2009learning} and CelebA \cite{liu2015faceattributes} for gender classification.
The CelebA images are center cropped to dimension $64\times 64$, and we randomly sample $5,000$ images for each class, which is same as CIFAR-10, while
FashionMNIST contains $6,000$ images for each class.
We adopt the same $3$-layer convolutional neural networks used in \cite{DC_distribution_matching} and \cite{KIP_neurips21} as the feature extractor.

\noindent\textbf{DC Settings.}
One important hyperparameter of DSA, DM and KIP is the ratio of image per class $r_{ipc}=\frac{\abs{\mathcal{S}}}{\abs{\mathcal{T}}}$.
We evaluate $r_{ipc}=0.002, 0.01$ for all methods, and for DM we add an extra evaluation 
$r_{ipc}=0.02$ due to its high efficiency on producing large synthetic set.
Note that $r_{ipc}$ influences the model training efficiency: the lower $r_{ipc}$, the faster model training.
We also consider ZCA preprocessing for KIP as it is reported to be effective for KIP performance improvement.
Appendix \ref{appendix:settings} contains more DC implementation details.

\noindent\textbf{Baselines.}
As for non-private baseline, we adopt subset sampled from $\mathcal{T}$ (this baseline is termed real data) and data generated by conditional GAN (cGAN or GAN for short) \cite{mirza2014conditional} which is trained on $\mathcal{T}$.
For private baseline, we choose DP-generators including GS-WGAN \cite{neurips20chen}, DP-MERF \cite{harder2021dp} and DP-Sinkhorn \cite{cao2021don}.
We compare the DC methods with baselines in terms of privacy and efficiency.

\noindent\textbf{MIA Settings \& Attack Metrics.}
For each dataset, we randomly split it into two subsets of equal amount of samples and choose one subset as $\mathcal{T}$ (member data).
We then synthesize dataset $\mathcal{S}$ on  $\mathcal{T}$, and train a model $f_\mathcal{S}$ (victim model) on $\mathcal{S}$.
The other subset becomes the non-member data used for testing the MIA performance.
The above process is called \textit{preparation of synthetic dataset}.
 
For loss-based MIA, we repeat the preparation of synthetic dataset $10$ times with different random seeds.
This gives us $10$ groups of $\mathcal{T}$, $\mathcal{S}$ and $f_\mathcal{S}$.
For each $f_\mathcal{S}$, we first select $N$ member samples from $\mathcal{T}$ and $N$ non-member samples, and choose an optimal threshold that maximizes the advantage score on the previously chosen $2N$ samples \cite{sablayrolles2019white}.
The threshold is then tested on another disjoint $2N$ samples composed by $N$ member samples and $N$ non-member samples to compute the advantage score of loss-based MIA.
We report the advantage (in percentage) defined as $2\times(acc-50\%)$ where $acc$ is the test accuracy of membership in percentage.

For LiRA \cite{carlini2021membership}, we repeat the preparation of synthetic dataset $N_m$ times with different random seeds, and obtain $N_m$ shadow $\mathcal{T}$, $\mathcal{S}$ and $f_\mathcal{S}$.
We set $N_m=256$ for DM and $N_m=64$ for KIP because of its lower training efficiency.
We omit DSA for LiRA due to longer training time. 
To attack a victim model, we compute the likelihoods of each sample with $N_m$ shadow $f_\mathcal{S}$ and determine the threshold of likelihood according to the requirements of false positive.
We use the Receiver Operating Characteristic (ROC) curve and Area Under Curve (AUC) score to evaluate the attack performance.
Remark that we adopt the strongest (and unrealistic) attack assumption (\ie, the attacker knows the membership), so that we investigate the privacy of DC-synthesized data under the \textit{worst} case.

\subsection{Membership Privacy of $f_\mathcal{S}$}
\label{subsec:membership_privacy_of_fS}

\begin{table}[htbp]
\centering
\caption{Advantage (\%) of loss-based MIA against models trained on real data (baseline) and data synthesized by DSA, DM and KIP with \textit{real data} initialization.
}
\label{tab:real_init_advantage}
\resizebox{\linewidth}{!}{
\begin{tabular}{ccccc}
\toprule
\textbf{Method} & $r_{ipc}$ & \textbf{FashionMNST} & \textbf{CIFAR-10} & \textbf{CelebA} \\ \midrule
\multirow{3}{*}{\begin{tabular}[c]{@{}c@{}}Real\\ (baseline,\\ non-private)\end{tabular}} & 0.002 & $46.67\pm 16.33$ & $72.00\pm 24.00$ & $100.00\pm 0.00$ \\ 
 & 0.01 & $21.00\pm 3.67$ & $92.80\pm 5.31$ & $84.00\pm 5.06$ \\ 
 & 0.02 & $17.33\pm 2.91$ & $82.60\pm 5.59$ & $77.00\pm 6.71$ \\ \midrule

\multirow{3}{*}{DM} & 0.002 & $78.17\pm 3.20$ & $49.80\pm 5.83$ & $37.00\pm 12.69$ \\ 
 & 0.01 & $83.67\pm 2.77$ & $64.20\pm 4.77$ & $47.00\pm 19.52$ \\ 
 & 0.02 & $83.00\pm 2.56$ & $68.20\pm 7.35$ & $53.00\pm 14.18$ \\ \midrule
 
\multirow{2}{*}{DSA} & 0.002 & $74.40\pm 2.65$ & $55.40\pm 8.20$ & $30.50\pm 8.16$ \\ 
 & 0.01 & $81.60\pm 2.27$ & $56.60\pm 2.95$ & $28.00\pm 3.74$ \\ \midrule

\multirow{2}{*}{\begin{tabular}[c]{@{}c@{}}KIP\\ (w/o ZCA)\end{tabular}} & 0.002 & $67.83\pm 4.54$ & $42.40\pm 4.80$ & $23.00\pm 11.87$ \\ 
 & 0.01 & $70.00\pm 2.47$ & $51.40\pm 5.73$ & $25.00\pm 15.65$ \\ \midrule
\multirow{2}{*}{\begin{tabular}[c]{@{}c@{}}KIP\\ (w/ ZCA)\end{tabular}} & 0.002 & $67.67\pm 4.42$ & $50.40\pm 5.35$ & $23.00\pm 15.52$ \\ 
 & 0.01 & $64.00\pm 4.23$ & $48.40\pm 6.62$ & $17.00\pm 18.47$ \\ \bottomrule
\end{tabular}}
\end{table}

\textbf{Real Data Initialization Leaks Membership Privacy.}
We begin with membership privacy leaked by $f_\mathcal{S}$ as mentioned in Remark \ref{remark:impact_init_on_privacy} of Proposition \ref{proposition:minimizer}.
We aim to verify that DC with real data initialization still leaks membership privacy of the data used for initialization.
Here, the data used for initialization are sampled from $\mathcal{T}$ during each time of preparation of synthetic dataset.
We launch loss-based MIA against $f_\mathcal{S}$ and adopt the real data baseline.

Table \ref{tab:real_init_advantage} shows the advantage score of loss-based MIA.
Here, we vary a little bit the attack setting: the advantage scores are computed with the data used for real initialization and the same amount of member data not used for initialization but involved in DC.
We observe that, on CIFAR-10 and CelebA, the synthetic dataset with real data initialization achieves lower advantage scores comparing to directly using real data for training (baseline).
This can be explained by Proposition \ref{proposition:minimizer}, which tells us that the synthesized data deviates slightly from the real data used for initialization.
However, on FashionMNIST, the baseline has lower advantage scores.
We suspect this is because FashionMNIST images are grey-scale and synthetic data contain more features that prone to be memorized by networks.
Loss distribution in Figure \ref{fig:real_init_hist} in Appendix \ref{appendix:loss_dist_dc_with_real_init} also demonstrates that \textit{synthetic data with real data initialization can still leak membership privacy}.

Next, we show that models trained on data synthesized by DC with random initialization are robust against both loss-based MIA and LiRA (the state-of-the-art MIA).

\textbf{MIA Robustness of Random Initialization.}
Because of random initialization, each member sample contributes equally to the synthetic dataset.
Thus, in this case, we follows the loss-base attack setting and set $N=1000$.
Table \ref{tab:loss_mia} provides the average and the standard variance of advantages for models trained on synthetic datasets by cGAN, DSA, DM and KIP.
The advantages are around $0$ for all $r_{ipc}$, signifying that the adversary cannot infer membership of $\mathcal{T}$.
Nevertheless, as long as the adversary has access to the generated images (which is included in our threat model), the membership of the GAN generator's training data (\ie, $\mathcal{T}$) can still be leaked (see Appendix \ref{appendix:mia_againt_gan}).
Meanwhile, as we show later in Figure \ref{fig:accuracy_barplot}, models trained on DC-synthesized data achieve higher accuracy scores than baseline (\ie, cGAN-generated data), demonstrating the higher utility of DC-synthesized data.

\begin{table}[htbp]
\centering
\caption{Advantage (\%) of loss-based MIA against models trained on data synthesized by cGAN (baseline), DSA, DM and KIP with \textit{random} initialization.}
\label{tab:loss_mia}
\resizebox{\linewidth}{!}{
\begin{tabular}{ccccc}
\toprule
\textbf{Methods} & $r_{ipc}$ & \textbf{FashionMNST} & \textbf{CIFAR-10} & \textbf{CelebA} \\ \midrule
\multirow{3}{*}{\begin{tabular}[c]{@{}c@{}}cGAN\\ (baseline,\\ non-private)\end{tabular}} & 0.002 & $0.29\pm 0.89$ & $-0.44\pm 1.88$ & $-0.57\pm 0.97$ \\ 
 & 0.01 & $0.18\pm 1.21$ & $-0.58\pm 2.09$ & $-0.81\pm 0.95$ \\ 
 & 0.02 & $0.04\pm 0.70$ & $-0.77\pm 1.59$ & $-0.47\pm 1.22$ \\ \midrule

\multirow{3}{*}{DM} & 0.002 & $-0.34\pm0.42$ & $0.31\pm 1.93$ & $-0.66\pm1.44$ \\ 
 & 0.01 & $-0.29\pm0.48$ & $1.06\pm 1.20$ & $-0.56\pm1.52$ \\ 
 & 0.02 & $0.18\pm0.53$ & $0.72\pm 0.70$ & $-0.67\pm1.18$ \\ \midrule
 
 \multirow{2}{*}{DSA} & 0.002 & $0.09\pm 0.51$ & $0.39\pm 1.04$ & $-0.39\pm 1.90$ \\ 
 & 0.01 & $0.52\pm 0.55$ & $1.27\pm 1.71$ & $-1.16\pm 0.90$ \\ \midrule
 
\multirow{2}{*}{\begin{tabular}[c]{@{}c@{}}KIP\\ (w/o ZCA)\end{tabular}} & 0.002 & $-1.13\pm1.84$ & $0.25\pm1.20$ & $-0.56\pm1.07$ \\ 
 & 0.01 & $-0.95\pm0.96$ & $0.25\pm1.80$ & $-1.51\pm0.69$ \\ \midrule
\multirow{2}{*}{\begin{tabular}[c]{@{}c@{}}KIP\\ (w/ ZCA)\end{tabular}} & 0.002 & $-0.56\pm2.02$ & $-0.64\pm1.86$ & $-1.06\pm1.10$ \\ 
 & 0.01 & $-1.69\pm1.96$ & $-0.22\pm1.27$ & $-1.80\pm1.91$ \\ \bottomrule
\end{tabular}}
\end{table}

\begin{figure*}
    \centering
    \resizebox{\linewidth}{!}{\includegraphics{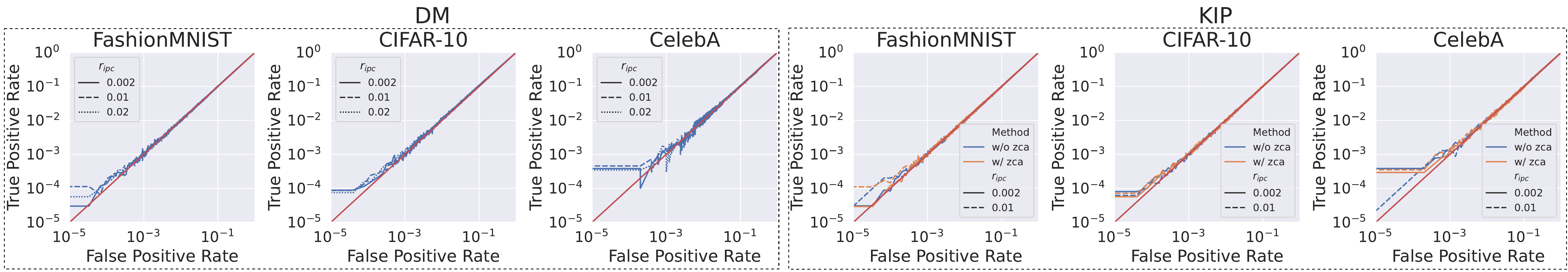}}
    \vspace*{-8mm}
    \caption{ROC curves of LiRA against models trained on data synthesized by DM (left three figures) and KIP (right three figures). 
    The solid, dashed and dotted lines stand for results of $r_{ipc}=0.002,0.01$ and $0.02$, respectively.
    In KIP figures, the orange and blue lines represent the results of KIP with and without ZCA preprocessing, respectively.
    The red diagonal represents random guess and the AUC scores of ROC curves are all under $0.51$.}
    \label{fig:LiRA_roc}
\end{figure*}

LiRA is a more powerful MIA because it can achieve higher TPR at low FPR \cite{carlini2021membership}, while the adversary's computational cost is higher.
Figure \ref{fig:LiRA_roc} provides the ROC curves of LiRA against $f_\mathcal{S}$.
We can observe that the ROC curves are close to the diagonal (red line) for all datasets and $r_{ipc}$.
The AUC scores of ROC curves are around $0.5$, indicating \textit{there is negligible attack benefit (low TPR) for the attacker compared with random guess}.
Recall that LiRA is evaluated on the whole dataset (half as member and other half as non-member), the minimum FPR value is $\frac{1}{5000}=2\times 10^{-4}$ for CelebA, $4\times10^{-5}$ for CIFAR-10 and $3.33\times10^{-5}$ for FashionMNIST.
Therefore, when FPR is close to $0$, the ROC curves have different shapes for different datasets.
However, we also notice that the TPR is close to FPR when FPR is around the minimum (\eg, FPR$\sim10^{-4}$ for CelebA), demonstrating that \textit{models trained on data synthesized by DC with random initialization is robust to LiRA at low FPR}.


\subsection{Comparison with Different Generators}
\label{subsec:compare_generator}

\begin{figure}
    \centering
    \resizebox{\linewidth}{!}{\includegraphics{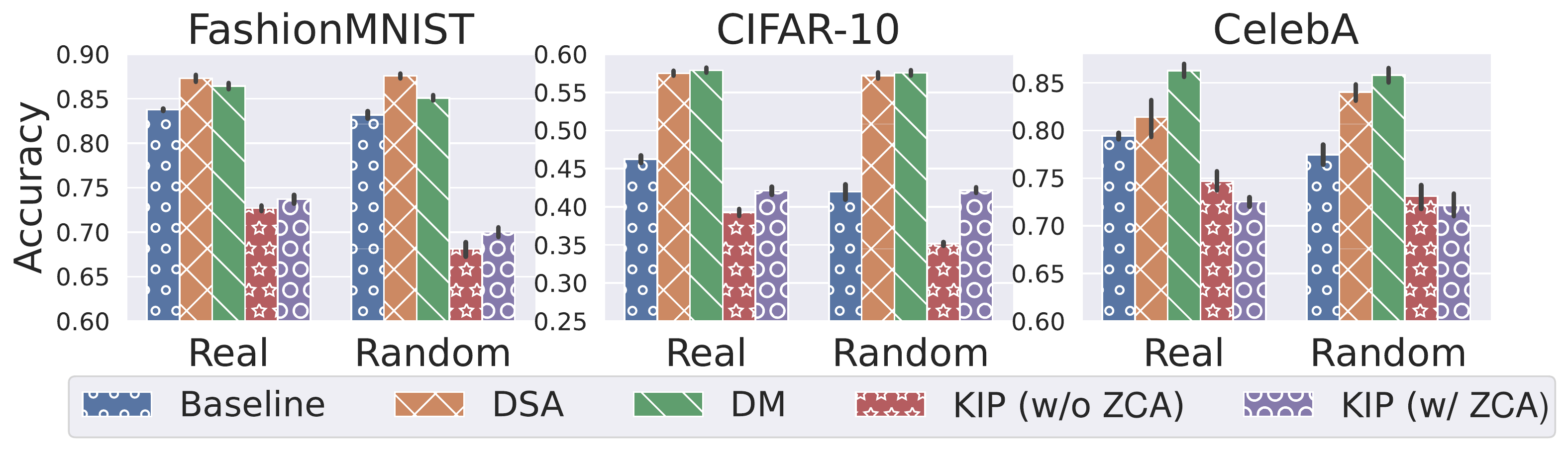}}
    \vspace*{-8mm}
    \caption{Accuracy of models trained on data synthesized by DSA, DM, KIP and on data generated by baselines for $r_{ipc}=0.01$. The x-axis represents initialization strategy. For real data and random initialization, the baselines are real data and cGAN-generated data, respectively.}
    \label{fig:accuracy_barplot}
\end{figure}

\begin{figure}
    \centering
    \resizebox{\linewidth}{!}{\includegraphics{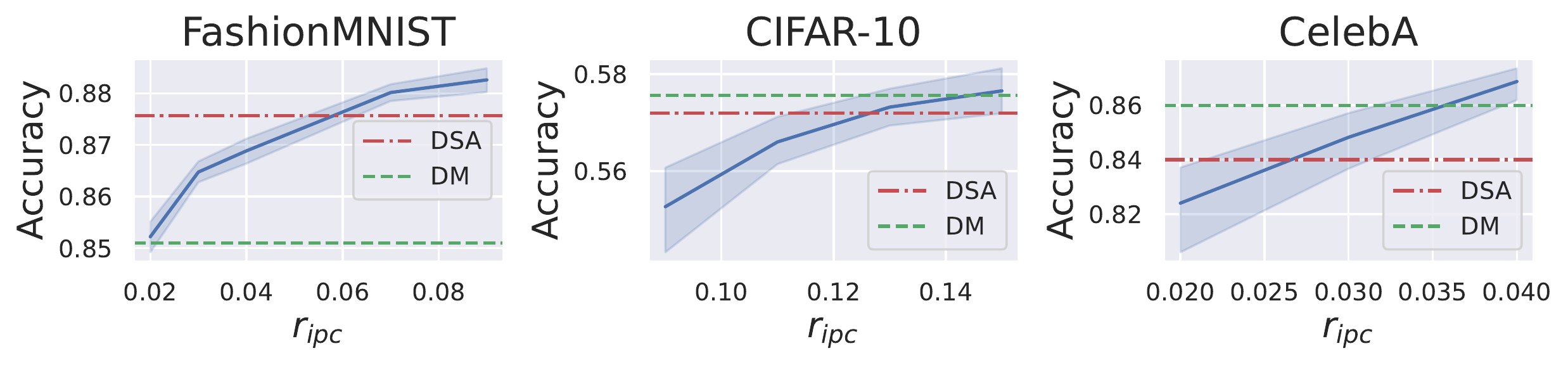}}
    \vspace*{-9mm}
    \caption{Accuracy of models trained on cGAN-generated data for different $r_{ipc}$. The horizontal lines are accuracy of models trained on data synthesized by DM (green, dashed) and DSA (red, dash-dotted) for $r_{ipc}=0.01$.}
    \label{fig:compare_efficiency}
\end{figure}

\begin{table}[htbp]
\centering
\caption{Utility comparison of dataset synthesized by DP-generators, DM and KIP. The utility is measured by the accuracy (\%) of models trained on the synthetic dataset. The results are estimated on FashionMNIST.}
\label{tab:compare_dp_gen}
\resizebox{\linewidth}{!}{
\begin{threeparttable}
\begin{tabular}{ccccc}
\toprule
\multirow{2}{*}{\textbf{Method}} & \multirow{2}{*}{\textbf{DP Budget}} & \multicolumn{3}{c}{$r_{ipc}$} \\ \cline{3-5} 
 &  & \multicolumn{1}{c}{$0.002$} & \multicolumn{1}{c}{$0.01$} & $0.02$ \\ \midrule
GS-WGAN & $\epsilon=10$ & \multicolumn{1}{c}{$53.53 \pm 0.42$} & \multicolumn{1}{c}{$51.85\pm 0.54$} & $50.10\pm 0.32$ \\ \midrule
\multirow{2}{*}{DP-MERF} 
& $\epsilon=10$ & \multicolumn{1}{c}{$52.18\pm 0.37$} & \multicolumn{1}{c}{$52.88\pm0.75$} & $50.73\pm 0.66$ \\ 
 & $\epsilon=2$ & \multicolumn{1}{c}{$60.41\pm0.78$} & \multicolumn{1}{c}{$55.14\pm0.61$} & $56.39\pm 0.45$ \\ \midrule
DP-Sinkhorn & $\epsilon=10$ & - & - & $70.9^\ast$ \\ \midrule
KIP (w/o ZCA) & $\hat{\epsilon}=1.25$ & \multicolumn{1}{c}{$73.70 \pm 1.13$} & \multicolumn{1}{c}{$68.11 \pm 1.33$} & - \\ 
KIP (w/ ZCA) & $\hat{\epsilon}=2.07$ & \multicolumn{1}{c}{$74.37 \pm 0.96$} & \multicolumn{1}{c}{$70.03 \pm 0.84$} & - \\ \midrule
DM & $\hat{\epsilon}=2.30$ & \multicolumn{1}{c}{$80.59 \pm 0.62$} & \multicolumn{1}{c}{$85.10 \pm 0.51$} & $86.13 \pm 0.34$ \\ \bottomrule
\end{tabular}
\begin{tablenotes} \footnotesize
\item $^{\ast}$ Results reported in the paper \cite{cao2021don} ($r_{ipc}=1$).
\end{tablenotes}
\end{threeparttable}
}
\end{table}
\textbf{GAN Generator.}
Figure \ref{fig:accuracy_barplot} compares the accuracy scores of models trained on synthetic datasets (DC-synthesized with random initialization under $r_{ipc}=0.01$).
We can find that under the same constraint of training efficiency (\ie, $r_{ipc}=0.01$), the DM and DSA outperform the other methods.
Note that models trained on KIP-synthesized data achieve lower accuracy than baseline because the loss is hard to converge for large $r_{ipc}$.
Nevertheless, for small $r_{ipc}$, the KIP significantly outperforms baselines on CIFAR-10 and CelebA (see Figure \ref{fig:accuracy_barplot0.002} in Appendix \ref{appendix:additional_exp}).

Then, we aim to know how DC improves model training efficiency compared to cGAN.
In other words, to achieve the same accuracy of $f_\mathcal{S}$, the difference between the $r_{ipc}$ that DC requires and the $r_{ipc}$ that cGAN requires can be seen as the model training efficiency that DC improves.
For different $r_{ipc}$ (the x-axis), Figure \ref{fig:compare_efficiency} shows the accuracy of models trained on cGAN-generated dataset whose ratio is $r_{ipc}$ (the blue solid curve).
The red and green horizontal lines represent the accuracy of $f_{\mathcal{S}}$ trained on dataset synthesized by DSA and DM for $r_{ipc}=0.01$, respectively.
We omit KIP here because of its lower utility than baselines.
Therefore, the $r_{ipc}$ of the intersection point of the red (resp. green) line and the blue curve is the $r_{ipc}$ of cGANs-generated dataset on which the models can be trained to achieve the same accuracy as DSA (resp. DM).
We can see that, cGAN needs to generate more data to train a model that achieves the same accuracy as models trained on data synthesized by DM and DSA, because the $r_{ipc}$ values indicated in the x-axis are all higher than $0.01$.
It is worth noting that DC improves the training efficiency (measured by $r_{ipc}$) by at least $2$ times than cGAN for $r_{ipc}=0.01$, because on FashionMNIST (the leftmost sub-figure in Figure \ref{fig:compare_efficiency})), cGAN requires to generate synthetic dataset of $r_{ipc}=0.02$ to achieve the same accuracy ($0.85$) as the DM-synthesized dataset ($r_{ipc}=0.01$).

\textbf{DP-generators.}
We estimate an empirical $\hat{\epsilon}$ based on the ratio of TPR and FPR computed by LiRA.
In Table \ref{tab:compare_dp_gen}, we compare the accuracy of models trained on DC-synthesized data and on data generated by recent DP-generators.
We reproduce DP-MERF and GS-WGAN according to the official implementation and adopt the reported results of DP-Sinkhorn.
We observe that the accuracy of models trained on data generated by the state-of-the-art DP-generator (DP-Sinkhorn) is still lower than DM-synthesized images, even the ratio for DP-Sinkhorn is $r_{ipc}=1$.
The reason is that DP is designed to defend against the strongest adversary who has access to the training process of generator.
Hence, data generated by DP-generators are of lower utility for model training because of the too strong defense requirement.

\subsection{Visual Privacy}
\label{subsec:vision_privacy}

\begin{figure}[t]
    \centering
    \resizebox{0.95\linewidth}{!}{\includegraphics{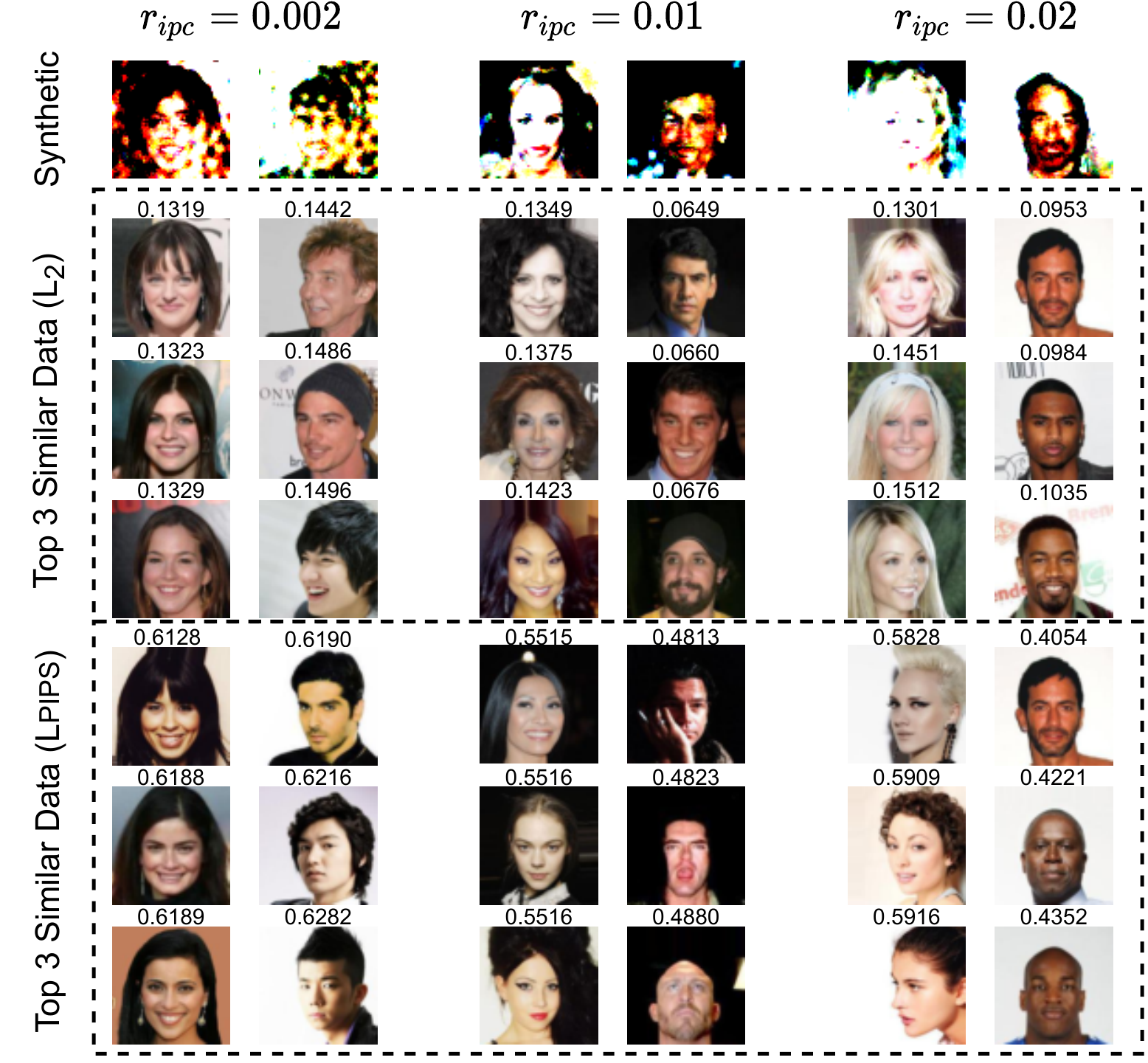}}
    \vspace*{-3mm}
    \caption{Examples of facial images that are most similar to synthetic data generated by DM with \textit{random} initialization. The value above each image is the distance ($L_2$ and LPIPS) between the image and the synthetic data (first row).
    Lower distance indicates higher similarity. Even though these real images have similar face contour, blurred facial details (\eg, eyes, nose) make it difficult for the adversary to infer the membership.}
    \label{fig:visual_privacy_example}
\end{figure}
The adversary can directly visualize synthetic data and compare with the target sample to infer the membership.
We visualize the synthetic images and use $L_2$ distance as well as the perceptual metric LPIPS \cite{zhang2018perceptual} with VGG backbone to measure the similarity between synthetic and real images.
Figure \ref{fig:visual_privacy_example} shows examples of DM-generated images and the their (top $3$) most similar real images, \ie, images of lowest $L_2$ and LPIPS distance with the synthetic image on the top of the column.
We observe that the real data share similar facial contour patterns with the synthetic images, but more fine-grained features, \eg, eye shape, are different, which explains why models trained on synthetic dataset protect the membership privacy of original data.
This can also explain why current MIAs fail on models trained on synthetic datasets: 
the generated synthetic training data have lost the private properties of real data and thus the adversary are not able to infer the privacy from models trained on such synthetic data.
\section{Discussion and Conclusion}

In this work, we make the first effort to introduce the emerging dataset condensation techniques into the privacy community and provide systematical audit including theoretical analysis of the privacy property and empirical evaluation, \ie visual privacy examination and robustness against loss-based MIA and LiRA on FashionMNIST, CIFAR-10 and CelebA datasets.

Our future work will attempt to generalize the theoretical findings to other DC methods.
This can be studied from the perspective of information loss (\eg data compression ratio).
Moreover, DC methods that satisfy formal DP formulation, \eg, $(\alpha, \epsilon)$-R\'{e}nyi DP \cite{DBLP:conf/csfw/Mironov17}, are worth exploring.

The current efforts of DC mainly focus on image classification, thus another interesting direction is the extension of the privacy benefit brought by DC to more complicated vision tasks (\eg, object detection) and non-vision tasks (\eg text and graph related applications).
In essence, DC methods can generalize to other machine learning tasks, as they learn the synthetic data by summarizing the distribution or discriminative information of real training data. Hence, their privacy advantage should also generalize to other tasks. 


\section*{Acknowledgements}
We would like to thank He Tong for the help in analyzing non-linear models and
the anonymous reviewers for constructive feedback.

\bibliography{ref}

\begin{thebibliography}{45}
\providecommand{\natexlab}[1]{#1}
\providecommand{\url}[1]{\texttt{#1}}
\expandafter\ifx\csname urlstyle\endcsname\relax
  \providecommand{\doi}[1]{doi: #1}\else
  \providecommand{\doi}{doi: \begingroup \urlstyle{rm}\Url}\fi

\bibitem[Bohdal et~al.(2020)Bohdal, Yang, and Hospedales]{bohdal2020flexible}
Bohdal, O., Yang, Y., and Hospedales, T.
\newblock Flexible dataset distillation: Learn labels instead of images.
\newblock \emph{NeurIPS Workshop}, 2020.

\bibitem[Cao et~al.(2021)Cao, Bie, Vahdat, Fidler, and Kreis]{cao2021don}
Cao, T., Bie, A., Vahdat, A., Fidler, S., and Kreis, K.
\newblock Don’t generate me: Training differentially private generative
  models with sinkhorn divergence.
\newblock \emph{Advances in Neural Information Processing Systems}, 34, 2021.

\bibitem[Carlini et~al.(2022)Carlini, Chien, Nasr, Song, Terzis, and
  Tramer]{carlini2021membership}
Carlini, N., Chien, S., Nasr, M., Song, S., Terzis, A., and Tramer, F.
\newblock Membership inference attacks from first principles.
\newblock In \emph{43rd {IEEE} Symposium on Security and Privacy, {SP} 2022}.
  {IEEE}, 2022.

\bibitem[Cazenavette et~al.(2022)Cazenavette, Wang, Torralba, Efros, and
  Zhu]{cazenavette2022distillation}
Cazenavette, G., Wang, T., Torralba, A., Efros, A.~A., and Zhu, J.-Y.
\newblock Dataset distillation by matching training trajectories.
\newblock In \emph{Proceedings of the IEEE/CVF Conference on Computer Vision
  and Pattern Recognition}, 2022.

\bibitem[Chen et~al.(2020{\natexlab{a}})Chen, Orekondy, and
  Fritz]{neurips20chen}
Chen, D., Orekondy, T., and Fritz, M.
\newblock Gs-wgan: A gradient-sanitized approach for learning differentially
  private generators.
\newblock In \emph{Neural Information Processing Systems (NeurIPS)},
  2020{\natexlab{a}}.

\bibitem[Chen et~al.(2020{\natexlab{b}})Chen, Yu, Zhang, and
  Fritz]{DBLP:conf/ccs/ChenYZF20}
Chen, D., Yu, N., Zhang, Y., and Fritz, M.
\newblock Gan-leaks: {A} taxonomy of membership inference attacks against
  generative models.
\newblock In \emph{{CCS} '20: 2020 {ACM} {SIGSAC} Conference on Computer and
  Communications Security}, pp.\  343--362, 2020{\natexlab{b}}.

\bibitem[Choquette-Choo et~al.(2021)Choquette-Choo, Tramer, Carlini, and
  Papernot]{pmlr-v139-choquette-choo21a}
Choquette-Choo, C.~A., Tramer, F., Carlini, N., and Papernot, N.
\newblock Label-only membership inference attacks.
\newblock In \emph{Proceedings of the 38th International Conference on Machine
  Learning}, volume 139 of \emph{Proceedings of Machine Learning Research},
  pp.\  1964--1974. PMLR, 2021.

\bibitem[Dwork et~al.(2006)Dwork, McSherry, Nissim, and
  Smith]{dwork2006calibrating}
Dwork, C., McSherry, F., Nissim, K., and Smith, A.
\newblock Calibrating noise to sensitivity in private data analysis.
\newblock In \emph{Theory of cryptography conference}, pp.\  265--284, 2006.

\bibitem[Estrach et~al.(2014)Estrach, Szlam, and LeCun]{estrach2014signal}
Estrach, J.~B., Szlam, A., and LeCun, Y.
\newblock Signal recovery from pooling representations.
\newblock In \emph{International conference on machine learning}, pp.\
  307--315. PMLR, 2014.

\bibitem[Fredrikson et~al.(2015)Fredrikson, Jha, and
  Ristenpart]{fredrikson2015model}
Fredrikson, M., Jha, S., and Ristenpart, T.
\newblock Model inversion attacks that exploit confidence information and basic
  countermeasures.
\newblock In \emph{CCS}, pp.\  1322--1333, 2015.

\bibitem[Giryes et~al.(2016)Giryes, Sapiro, and Bronstein]{giryes2016deep}
Giryes, R., Sapiro, G., and Bronstein, A.~M.
\newblock Deep neural networks with random gaussian weights: A universal
  classification strategy?
\newblock \emph{IEEE Transactions on Signal Processing}, 64\penalty0
  (13):\penalty0 3444--3457, 2016.

\bibitem[Goodfellow et~al.(2014)Goodfellow, Pouget-Abadie, Mirza, Xu,
  Warde-Farley, Ozair, Courville, and Bengio]{goodfellow2014generative}
Goodfellow, I., Pouget-Abadie, J., Mirza, M., Xu, B., Warde-Farley, D., Ozair,
  S., Courville, A., and Bengio, Y.
\newblock Generative adversarial nets.
\newblock \emph{Advances in neural information processing systems}, 27, 2014.

\bibitem[Harder et~al.(2021)Harder, Adamczewski, and Park]{harder2021dp}
Harder, F., Adamczewski, K., and Park, M.
\newblock Dp-merf: Differentially private mean embeddings with randomfeatures
  for practical privacy-preserving data generation.
\newblock In \emph{International Conference on Artificial Intelligence and
  Statistics}, pp.\  1819--1827, 2021.

\bibitem[Hinton et~al.(2015)Hinton, Vinyals, and Dean]{hinton2015distilling}
Hinton, G., Vinyals, O., and Dean, J.
\newblock Distilling the knowledge in a neural network.
\newblock \emph{arXiv preprint arXiv:1503.02531}, 2015.

\bibitem[Jacot et~al.(2018)Jacot, Hongler, and Gabriel]{neural_tangent_kernel}
Jacot, A., Hongler, C., and Gabriel, F.
\newblock Neural tangent kernel: Convergence and generalization in neural
  networks.
\newblock In \emph{Advances in Neural Information Processing Systems 31: Annual
  Conference on Neural Information Processing Systems 2018, NeurIPS 2018}, pp.\
   8580--8589, 2018.

\bibitem[Jin et~al.(2022)Jin, Zhao, Zhang, Liu, Tang, and Shah]{jin2022graph}
Jin, W., Zhao, L., Zhang, S., Liu, Y., Tang, J., and Shah, N.
\newblock Graph condensation for graph neural networks.
\newblock \emph{ICLR}, 2022.

\bibitem[Kairouz et~al.(2015)Kairouz, Oh, and
  Viswanath]{kairouz2015composition}
Kairouz, P., Oh, S., and Viswanath, P.
\newblock The composition theorem for differential privacy.
\newblock In \emph{International conference on machine learning}, pp.\
  1376--1385. PMLR, 2015.

\bibitem[Krizhevsky et~al.(2009)Krizhevsky, Hinton,
  et~al.]{krizhevsky2009learning}
Krizhevsky, A., Hinton, G., et~al.
\newblock Learning multiple layers of features from tiny images.
\newblock 2009.

\bibitem[LeCun et~al.(1998)LeCun, Bottou, Bengio, Haffner,
  et~al.]{lecun1998gradient}
LeCun, Y., Bottou, L., Bengio, Y., Haffner, P., et~al.
\newblock Gradient-based learning applied to document recognition.
\newblock \emph{Proceedings of the IEEE}, 86\penalty0 (11):\penalty0
  2278--2324, 1998.

\bibitem[Li \& Zhang(2021)Li and Zhang]{DBLP:conf/ccs/LiZ21}
Li, Z. and Zhang, Y.
\newblock Membership leakage in label-only exposures.
\newblock In \emph{{CCS} '21: 2021 {ACM} {SIGSAC} Conference on Computer and
  Communications Security}, pp.\  880--895, 2021.

\bibitem[Liu et~al.(2015)Liu, Luo, Wang, and Tang]{liu2015faceattributes}
Liu, Z., Luo, P., Wang, X., and Tang, X.
\newblock Deep learning face attributes in the wild.
\newblock In \emph{Proceedings of International Conference on Computer Vision
  (ICCV)}, December 2015.

\bibitem[Lyu et~al.(2020)Lyu, Yu, Zhao, and Yang]{lyu2020threats}
Lyu, L., Yu, H., Zhao, J., and Yang, Q.
\newblock Threats to federated learning.
\newblock In \emph{Federated Learning}, pp.\  3--16. Springer, 2020.

\bibitem[Melis et~al.(2019)Melis, Song, De~Cristofaro, and
  Shmatikov]{melis2019exploiting}
Melis, L., Song, C., De~Cristofaro, E., and Shmatikov, V.
\newblock Exploiting unintended feature leakage in collaborative learning.
\newblock In \emph{SP}, pp.\  691--706, 2019.

\bibitem[Mironov(2017)]{DBLP:conf/csfw/Mironov17}
Mironov, I.
\newblock R{\'{e}}nyi differential privacy.
\newblock In \emph{30th {IEEE} Computer Security Foundations Symposium, {CSF}
  2017, Santa Barbara, CA, USA, August 21-25, 2017}, pp.\  263--275. {IEEE}
  Computer Society, 2017.

\bibitem[Mirza \& Osindero(2014)Mirza and Osindero]{mirza2014conditional}
Mirza, M. and Osindero, S.
\newblock Conditional generative adversarial nets.
\newblock \emph{arXiv preprint arXiv:1411.1784}, 2014.

\bibitem[Nasr et~al.(2021)Nasr, Songi, Thakurta, Papemoti, and
  Carlin]{nasr2021adversary}
Nasr, M., Songi, S., Thakurta, A., Papemoti, N., and Carlin, N.
\newblock Adversary instantiation: Lower bounds for differentially private
  machine learning.
\newblock In \emph{2021 IEEE Symposium on Security and Privacy (SP)}, pp.\
  866--882. IEEE, 2021.

\bibitem[Nguyen et~al.(2021{\natexlab{a}})Nguyen, Chen, and Lee]{KIP_iclr21}
Nguyen, T., Chen, Z., and Lee, J.
\newblock Dataset meta-learning from kernel ridge-regression.
\newblock In \emph{International Conference on Learning Representations},
  2021{\natexlab{a}}.

\bibitem[Nguyen et~al.(2021{\natexlab{b}})Nguyen, Novak, Xiao, and
  Lee]{KIP_neurips21}
Nguyen, T., Novak, R., Xiao, L., and Lee, J.
\newblock Dataset distillation with infinitely wide convolutional networks.
\newblock In \emph{Thirty-Fifth Conference on Neural Information Processing
  Systems}, 2021{\natexlab{b}}.

\bibitem[Powell(1964)]{DBLP:journals/cj/Powell64}
Powell, M. J.~D.
\newblock An efficient method for finding the minimum of a function of several
  variables without calculating derivatives.
\newblock \emph{Comput. J.}, 7\penalty0 (2):\penalty0 155--162, 1964.

\bibitem[Rezaei \& Liu(2021)Rezaei and Liu]{rezaei2021difficulty}
Rezaei, S. and Liu, X.
\newblock On the difficulty of membership inference attacks.
\newblock In \emph{Proceedings of the IEEE/CVF Conference on Computer Vision
  and Pattern Recognition}, pp.\  7892--7900, 2021.

\bibitem[Sablayrolles et~al.(2019)Sablayrolles, Douze, Schmid, Ollivier, and
  J{\'e}gou]{sablayrolles2019white}
Sablayrolles, A., Douze, M., Schmid, C., Ollivier, Y., and J{\'e}gou, H.
\newblock White-box vs black-box: Bayes optimal strategies for membership
  inference.
\newblock In \emph{International Conference on Machine Learning}, pp.\
  5558--5567, 2019.

\bibitem[Shokri et~al.(2017)Shokri, Stronati, Song, and
  Shmatikov]{shokri2017membership}
Shokri, R., Stronati, M., Song, C., and Shmatikov, V.
\newblock Membership inference attacks against machine learning models.
\newblock In \emph{2017 IEEE Symposium on Security and Privacy (SP)}, pp.\
  3--18, 2017.

\bibitem[Such et~al.(2020)Such, Rawal, Lehman, Stanley, and
  Clune]{such2020generative}
Such, F.~P., Rawal, A., Lehman, J., Stanley, K.~O., and Clune, J.
\newblock Generative teaching networks: Accelerating neural architecture search
  by learning to generate synthetic training data.
\newblock \emph{ICML}, 2020.

\bibitem[Sucholutsky \& Schonlau(2019)Sucholutsky and
  Schonlau]{sucholutsky2019soft}
Sucholutsky, I. and Schonlau, M.
\newblock Soft-label dataset distillation and text dataset distillation.
\newblock \emph{arXiv preprint arXiv:1910.02551}, 2019.

\bibitem[Wang et~al.(2022)Wang, Zhao, Peng, Zhu, Yang, Wang, Huang, Bilen,
  Wang, and You]{wang2022cafe}
Wang, K., Zhao, B., Peng, X., Zhu, Z., Yang, S., Wang, S., Huang, G., Bilen,
  H., Wang, X., and You, Y.
\newblock Cafe: Learning to condense dataset by aligning features.
\newblock \emph{Proceedings of the IEEE/CVF Conference on Computer Vision and
  Pattern Recognition}, 2022.

\bibitem[Wang et~al.(2018)Wang, Zhu, Torralba, and Efros]{dataset_distillation}
Wang, T., Zhu, J., Torralba, A., and Efros, A.~A.
\newblock Dataset distillation.
\newblock \emph{CoRR}, abs/1811.10959, 2018.
\newblock URL \url{http://arxiv.org/abs/1811.10959}.

\bibitem[Wang et~al.(2021)Wang, Ding, Xiao, Kifer, and Zhang]{wang2021dpgen}
Wang, Y., Ding, Z., Xiao, Y., Kifer, D., and Zhang, D.
\newblock Dpgen: Automated program synthesis for differential privacy.
\newblock In \emph{Proceedings of the 2021 ACM SIGSAC Conference on Computer
  and Communications Security}, pp.\  393--411, 2021.

\bibitem[Webster et~al.(2021)Webster, Rabin, Simon, and
  Jurie]{webster2021person}
Webster, R., Rabin, J., Simon, L., and Jurie, F.
\newblock This person (probably) exists. identity membership attacks against
  gan generated faces.
\newblock \emph{arXiv preprint arXiv:2107.06018}, 2021.

\bibitem[Xiao et~al.(2017)Xiao, Rasul, and Vollgraf]{xiao2017fashion}
Xiao, H., Rasul, K., and Vollgraf, R.
\newblock Fashion-mnist: a novel image dataset for benchmarking machine
  learning algorithms.
\newblock \emph{arXiv preprint arXiv:1708.07747}, 2017.

\bibitem[Xie et~al.(2018)Xie, Lin, Wang, Wang, and Zhou]{xie2018differentially}
Xie, L., Lin, K., Wang, S., Wang, F., and Zhou, J.
\newblock Differentially private generative adversarial network.
\newblock \emph{arXiv preprint arXiv:1802.06739}, 2018.

\bibitem[Yu et~al.(2021)Yu, Zhang, Chen, Yin, and Liu]{yu2021does}
Yu, D., Zhang, H., Chen, W., Yin, J., and Liu, T.-Y.
\newblock How does data augmentation affect privacy in machine learning?
\newblock In \emph{Proceedings of the AAAI Conference on Artificial
  Intelligence}, volume~35, pp.\  10746--10753, 2021.

\bibitem[Zhang et~al.(2018)Zhang, Isola, Efros, Shechtman, and
  Wang]{zhang2018perceptual}
Zhang, R., Isola, P., Efros, A.~A., Shechtman, E., and Wang, O.
\newblock The unreasonable effectiveness of deep features as a perceptual
  metric.
\newblock In \emph{CVPR}, 2018.

\bibitem[Zhao \& Bilen(2021{\natexlab{a}})Zhao and
  Bilen]{DC_distribution_matching}
Zhao, B. and Bilen, H.
\newblock Dataset condensation with distribution matching.
\newblock \emph{CoRR}, abs/2110.04181, 2021{\natexlab{a}}.

\bibitem[Zhao \& Bilen(2021{\natexlab{b}})Zhao and Bilen]{DSA}
Zhao, B. and Bilen, H.
\newblock Dataset condensation with differentiable siamese augmentation.
\newblock In \emph{International Conference on Machine Learning},
  2021{\natexlab{b}}.

\bibitem[Zhao et~al.(2021)Zhao, Mopuri, and Bilen]{DC_gradient_matching}
Zhao, B., Mopuri, K.~R., and Bilen, H.
\newblock Dataset condensation with gradient matching.
\newblock In \emph{International Conference on Learning Representations}, 2021.

\end{thebibliography}
\bibliographystyle{icml2022}

\newpage
\onecolumn
\appendix

\section{Proof of Proposition \ref{proposition:DM_linear}}
\label{appendix:proof1}
We begin our analysis of DM with the linear extractor $\psi_{\boldsymbol\theta}:\mathbb{R}^d\rightarrow \mathbb{R}^k$ such that 
$k<d$, $\boldsymbol\theta=[\theta_{i,j}] \in \mathbb{R}^{k \times d}$ and for an input $\mathbf{x}$, $\psi_{\boldsymbol\theta}(\mathbf{x}) = \boldsymbol\theta \mathbf{x}$.
We also omit the differentiable Siamese augmentation to simplify the analysis.
As the representation extractors $\psi_{\boldsymbol\theta}$ in DM are randomly initialized, we assume that the extractor parameters follow the standard normal distribution and are identically and independently distributed (i.i.d), \ie, $\theta_{i,j}  \stackrel{iid}{\sim}  \mathcal{N}(0,1)$.
Thus, Equation \eqref{eq:loss_DM} becomes the expectation over $\norm{\mathbf{d}_{DM}}^2$ where $\mathbf{d}_{DM}$ is defined as:
\begin{equation}
   \label{eq:d_DM}
   \mathbf{d}_{DM} \coloneqq  \btheta(\frac{1}{\abs{\mathcal{T}}}\sum\limits_{i=1}^{\abs{\mathcal{T}}} \mathbf{x}_i - \frac{1}{\abs{\mathcal{S}}}\sum\limits_{i=1}^{\abs{\mathcal{S}}} \mathbf{s}_i).
\end{equation}
Hence, $L_{DM} = \mathbb{E}_{\boldsymbol\theta \sim \mathcal{N}(0,1)} \norm{\mathbf{d}_{DM}}^2$.
The optimization of $\mathcal{S}$ with SGD relies on the gradient of \eqref{eq:loss_DM}.
Given a sampled model parameter $\boldsymbol\theta$, for some synthetic sample $\mathbf{s}_j$, we have: 
\begin{equation}
 \begin{aligned}
    \frac{\partial \norm{\mathbf{d}_{DM}}^2}{\partial\mathbf{s}_i} &= -\frac{2}{\abs{\mathcal{S}}}(\mathbf{d}_{DM})^\top \boldsymbol\theta = -\frac{2}{\abs{\mathcal{S}}}(\frac{1}{\abs{\mathcal{T}}}\sum\limits_{j=1}^{\abs{\mathcal{T}}} \mathbf{x}_j - \frac{1}{\abs{\mathcal{S}}}\sum\limits_{j=1}^{\abs{\mathcal{S}}} \mathbf{s}_j)^\top \cdot \boldsymbol\theta^\top \boldsymbol\theta.
    \end{aligned}
\end{equation}
Hence, we obtain:
\begin{equation}
\label{eq:gradient_loss_DM}
\boxed{\frac{\partial L_{DM}}{\partial \mathbf{s}_i} = \frac{\partial \mathbb{E}_{\boldsymbol\theta} \norm{\mathbf{d}_{DM}}^2}{\partial \mathbf{s}_i} = \mathbb{E}_{\boldsymbol\theta} \frac{\partial \norm{\mathbf{d}_{DM}}^2}{\partial\mathbf{s}_i} = -\frac{2}{\abs{\mathcal{S}}}(\frac{1}{\abs{\mathcal{T}}}\sum\limits_{j=1}^{\abs{\mathcal{T}}} \mathbf{x}_j - \frac{1}{\abs{\mathcal{S}}}\sum\limits_{j=1}^{\abs{\mathcal{S}}} \mathbf{s}_j)^\top \cdot \mathbb{E}[\boldsymbol\theta^\top \boldsymbol\theta],}
\end{equation}
where $\mathbb{E}[\boldsymbol\theta^\top \boldsymbol\theta] = k\mathbf{I}_d$ by definition of $\boldsymbol\theta$, and $\mathbf{I}_d$ is the identity matrix of  $\mathbb{R}^d$.
Equation \eqref{eq:gradient_loss_DM} indicates that the optimization direction of synthetic sample $\mathbf{s}_i$ is the direction of moving barycenter of $\mathcal{S}$ to the barycenter of $\mathcal{T}$.
To conceptually interpret, the optimization of \eqref{eq:loss_DM} will move the initialized $\mathcal{S}$ until the barycenter coincides with that of $\mathcal{T}$ because of the existence of minimizer (Assumption \ref{assump:convergence}) where the left hand-side of \eqref{eq:gradient_loss_DM} should be $\mathbf{0}$.

\section{Proof of Proposition \ref{proposition:minimizer}}
\label{appendix:proof2}
\noindent\textbf{Case of real data initialization.}
Suppose that each $\mathbf{s}_i\in\mathcal{S}$ is sampled from $\mathcal{T}$ and
we can consider $\mathbf{s}_i = \mathbf{x}_i$ as initialization for simplicity.
According to \eqref{eq:gradient_loss_DM}, all $\mathbf{s}_i$ are optimized until the barycenters of $\mathcal{S}$ and $\mathcal{T}$ coincide.
Observe that each $\mathbf{s}_j^\ast\in \text{span}(\mathcal{T})$, because the projection components of $\text{span}(\mathcal{T})^\perp$ remain zero throughout the optimization process of DM.
Thus, one solution of minimizer elements $\mathbf{s}_i^\ast \in \mathcal{S}^\ast$ with the real data initialization can be:
\begin{equation}
    \boxed{\mathbf{s}_i^\ast = \mathbf{x}_i + \frac{1}{\abs{\mathcal{T}}}\sum\limits_{j=1}^{\abs{\mathcal{T}}}\mathbf{x}_j - \frac{1}{\abs{\mathcal{S}}}\sum\limits_{j=1}^{\abs{\mathcal{S}}}\mathbf{s}_j.}
\end{equation}
When $\abs{\mathcal{S}}$ and $\abs{\mathcal{T}}$ are large (\eg, $>50$), we can consider $\mathbf{s}_i^\ast \approx \mathbf{x}_i$, thus \textit{initialization with real data in DM still risks of membership privacy leakage}.

\noindent\textbf{Case of random initialization.}
The synthetic data are initialized as vectors of multivariate normal distribution, \ie, 
\begin{equation}
    \forall \mathbf{s}_i\in \mathcal{S}, \mathbf{s}_i \sim \mathcal{N}(\mathbf{0},\mathbf{I}_d).
\end{equation}
Each synthetic sample $\mathbf{s}_i$ can be written as a vector $[s_{i,1},\cdots,s_{i,d}]$ under the basis $\mathcal{E}$ where $\forall j, s_{i,j}\stackrel{i.i.d.}{\sim}\mathcal{N}(0,1)$, because $\mathbf{s}_i$'s covariance matrix remains identity matrix under any orthogonal transformation (\ie, orthogonal basis).
Thus, we can decompose $\mathbf{d}_{DM}$ to the projections on subspace $\text{span}(\mathcal{T})$ and its orthogonal complement.
Formally, we have
\begin{equation}
    \begin{aligned}
       \norm{\mathbf{d}_{DM}}^2 
       &= \norm{\boldsymbol\theta _{1:d_\mathcal{T}} \text{Proj}_{\mathcal{E}_\mathcal{T}}(\Delta_{\mathcal{S}, \mathcal{T}})}^2 +  \underbrace{\norm{\boldsymbol\theta _{d_\mathcal{T}:d} \text{Proj}_{\mathcal{E}_\mathcal{T}^\perp}(\Delta_{\mathcal{S}, \mathcal{T}})}^2}_{\norm{\mathbf{d}_{DM}}^2_{\mathcal{E}_\mathcal{T}^\perp}}
       \\ &+ 2<\boldsymbol\theta _{1:d_\mathcal{T}} \text{Proj}_{\mathcal{E}_\mathcal{T}}(\Delta_{\mathcal{S}, \mathcal{T}}), \boldsymbol\theta _{d_\mathcal{T}:d} \text{Proj}_{\mathcal{E}_\mathcal{T}^\perp}(\Delta_{\mathcal{S}, \mathcal{T}})>,
    \end{aligned}
\end{equation}
where $\boldsymbol\theta_{a:b}$ represents the submatrix composed by $a$-th column to the $b$-th column, $\text{Proj}_{V}$ is the projection operator onto subspace $V$ and 
\begin{equation}
    \Delta_{\mathcal{S}, \mathcal{T}} \coloneqq \frac{1}{\abs{\mathcal{T}}}\sum\limits_{i=1}^{\abs{\mathcal{T}}} \mathbf{x}_i - \frac{1}{\abs{\mathcal{S}}}\sum\limits_{i=1}^{\abs{\mathcal{S}}} \mathbf{s}_i.
\end{equation}
Note that 
\begin{equation}
    \text{Proj}_{\mathcal{E}_\mathcal{T}^\perp}(\Delta_{\mathcal{S}, \mathcal{T}}) = \frac{1}{\abs{\mathcal{S}}}\sum\limits_{i=1}^{\abs{\mathcal{S}}}\text{Proj}_{\mathcal{E}_\mathcal{T}^\perp}(\mathbf{s}_i),
\end{equation}
because $\mathbf{x}_i\in \text{span}(\mathcal{T})=\text{span}(\mathcal{E}_\mathcal{T})$ for each $\mathbf{x}_i$.
Let $\mathbf{s}_{i, \mathcal{E}_\mathcal{T}^\perp} =\text{Proj}_{\mathcal{E}_\mathcal{T}^\perp}(\mathbf{s}_i)$, then we have
\begin{equation}
    \mathbb{E}_{\boldsymbol\theta}\frac{\partial \norm{\mathbf{d}_{DM}}^2_{\mathcal{E}_\mathcal{T}^\perp}}{\partial \mathbf{s}_{i, \mathcal{E}_\mathcal{T}^\perp}} = \frac{2}{\abs{\mathcal{S}}^2}(\sum\limits_{j=1}^{\abs{\mathcal{S}}}\mathbf{s}_{j, \mathcal{E}_\mathcal{T}^\perp})^\top\mathbb{E}_{\boldsymbol\theta}[(\boldsymbol\theta_{d_\mathcal{T}:d})^\top \boldsymbol\theta _{d_\mathcal{T}:d}],
\end{equation}
because $\mathbb{E}_{\boldsymbol\theta}[\boldsymbol\theta_{1:d_\mathcal{T}}^\top \boldsymbol\theta_{d_\mathcal{T}:d}]=\mathbf{0}$.
Therefore, the expectation of the above equation is the optimization direction of the projection of $\mathbf{s}_j$ on the subspace $(\text{span}(\mathcal{T}))^\perp$:
\begin{equation}
\begin{aligned}
   \frac{\partial L_{DM}}{\partial \mathbf{s}_{i, \mathcal{E}_\mathcal{T}^\perp}} &=
    \mathbb{E}_{\boldsymbol\theta} \frac{\partial \norm{\mathbf{d}_{DM}}^2_{\mathcal{E}_\mathcal{T}^\perp}}{\partial \mathbf{s}_{i, \mathcal{E}_\mathcal{T}^\perp}} 
    = \frac{2\mathbb{E}[(\boldsymbol\theta_{d_\mathcal{T}:d})^\top \boldsymbol\theta _{d_\mathcal{T}:d}]}{\abs{\mathcal{S}}^2}(\sum\limits_{j=1}^{\abs{\mathcal{S}}}\mathbf{s}_{j, \mathcal{E}_\mathcal{T}^\perp})^\top.
    \end{aligned}
\end{equation}
Note that $\mathbb{E}[(\boldsymbol\theta_{d_\mathcal{T}:d})^\top \boldsymbol\theta _{d_\mathcal{T}:d}]=k\mathbf{I}_{d-d_{\mathcal{T}}}$, thus the optimization direction is aligned with the barycenter of all $\mathbf{s}_{i, \mathcal{E}_\mathcal{T}^\perp}$ and will converge to $0$ when
\begin{equation}
    \frac{1}{\abs{\mathcal{S}}}\sum\limits_{i=1}^{\abs{\mathcal{S}}}\mathbf{s}_{i, \mathcal{E}_\mathcal{T}^\perp}=\mathbf{0}_{ \mathcal{E}_\mathcal{T}^\perp}.
\end{equation}
Since the initialization of $\mathcal{S}$ is essentially noise of standard normal distribution, the empirical average of $\mathbf{s}_{i, \mathcal{E}_\mathcal{T}^\perp}$ is close to $\mathbf{0}$ (by law of large numbers), thus we can consider that the projection component of minimizer $\mathbf{s}_{i, \mathcal{E}_\mathcal{T}^\perp}^\ast$ is close to the initialized value, \ie,
\begin{equation}
    \boxed{\forall \mathbf{s}_{i, \mathcal{E}_\mathcal{T}^\perp}^\ast\in\mathcal{S}^\ast, \mathbf{s}_{i, \mathcal{E}_\mathcal{T}^\perp}^\ast\approx \mathbf{s}_{i, \mathcal{E}_\mathcal{T}^\perp}.}
\end{equation}

Similar as the case of real data initialization, the projection components on $\text{span}(\mathcal{T})$ of $\mathbf{s}_i$ are optimized to verify the first property of Proposition \ref{proposition:DM_linear}, \ie, the projection component of $i$-th minimizer $\mathbf{s}_{i, \mathcal{E}_\mathcal{T}}^\ast$ becomes
\begin{equation}
    \boxed{\mathbf{s}_{i, \mathcal{E}_\mathcal{T}}^\ast = \mathbf{s}_{i, \mathcal{E}_\mathcal{T}} + \frac{1}{\abs{\mathcal{T}}}\sum\limits_{j=1}^{\abs{\mathcal{T}}}\mathbf{x}_j - \frac{1}{\abs{\mathcal{S}}}\sum\limits_{j=1}^{\abs{\mathcal{S}}}\mathbf{s}_{j, \mathcal{E}_\mathcal{T}}.}
\end{equation}

\subsection{Empirical verification}
\label{appendix_subsec:emp_verify_minimizer}

\begin{figure}[htbp]
    \centering
    \resizebox{0.9\linewidth}{!}{\includegraphics{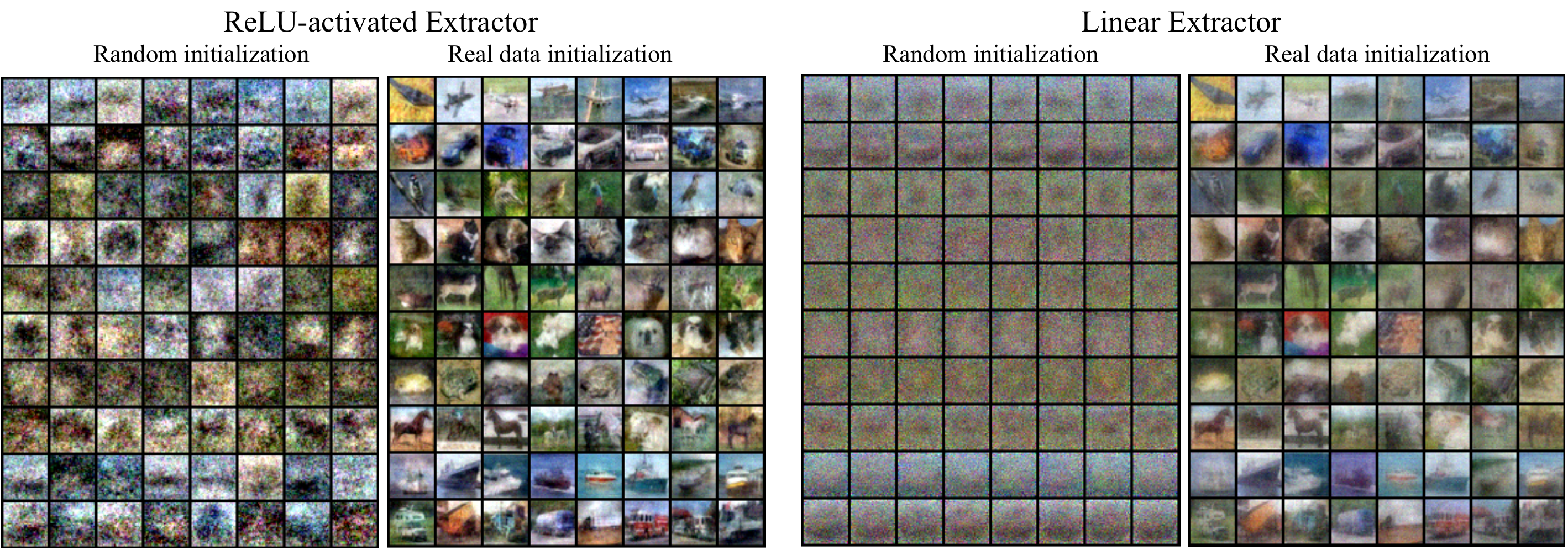}}
    \caption{Synthetic images with random noise/real data initialization and with linear/ReLU-activated extractor of CIFAR-10.}
    \label{fig:verify_proposition_minimizer}
\end{figure}

We empirically verify our conclusions for random and real data initializations in Figure~\ref{fig:verify_proposition_minimizer}.
The images are synthesized from CIFAR-10 by DM using linear extractor of embedding dimension $2048$, and each line contains images from the same class.
On the right side, we plot the images synthesized with random initialization and real data initialization.
We can observe that images synthesized with random initialization resemble combination of noise and class-dependent background, which verifies our conclusion of random initialization: synthetic data with random initialization are composed of barycenter of original data in space $\text{span}$ and initialized noise in space $\text{span}(\mathcal{T})^\perp$ (see \eqref{eq:minimizer_random}).
Note that even in this case, models trained on synthetic data can still achieve validation accuracy around 27\%.

On the other hand, real data initialization generates little changes on the images used for initialization, which verifies the conclusion of real data initialization: synthetic data with real data initialization are composed of images used for initialization and the barycenter distance vector (see \eqref{eq:minimizer_real_data}).

Besides linear extractor, we also investigate the impact of activation function.
On the left of Figure~\ref{fig:verify_proposition_minimizer}, we show images synthesized by DM using ReLU-activated extractor (ReLU on top of linear extractor).
We can see that the existence of ReLU results in better convergence of DC and thus better image quality for both random and real data initialization.
A potential reason is that ReLU changes the DC optimization and can lead to different local minima other than that found by using linear extractor.
That is, data synthesized in this case are possibly composed by barycenter of a certain group of similar images (\eg, images of brown horse heading towards right with grass background) within the same class and a orthogonal noise vector.
For example, CIFAR-10 synthetic images of class ``horse'' (third last line of leftmost figure in Figure~\ref{fig:verify_proposition_minimizer}) are noisy but contain different backgrounds which should be the barycenter of different image group of class ``horse'': there are numerous horse images in CIFAR-10 where the horse head towards the right or left.
This observation also confirms that ReLU improves generalization of neural networks.
Appendix~\ref{appendix:non_linear_extractor} encompasses more detailed analysis for non-linear extractor.

\section{Proof of Proposition \ref{proposition:KL_div}}
\label{appendix:proof3}

We aim to quantify the membership privacy leakage of a member $\mathbf{x}$ with the Kullback-Leibler (KL) divergence of model parameter distributions.
Without loss of generality, we study how the last element $\mathbf{x}_{\abs{\mathcal{T}}}$ influences the model parameter distribution.
Let $\mathcal{T}^\prime$ denote $\mathcal{T}\setminus\{\mathbf{x}_{\abs{\mathcal{T}}}\}$, where $\mathcal{T}=\{\mathbf{x}_1,\cdots, \mathbf{x}_{\abs{\mathcal{T}}}\}$.
The synthetic datasets by DM based on $\mathcal{T}$ and $\mathcal{T}^\prime$ are noted as $\mathcal{S}$ and $\mathcal{S}^\prime$, respectively, and $\abs{\mathcal{S}}=\abs{\mathcal{S}^\prime}$.
In addition, we denote $p(\boldsymbol\theta)=\mathbb{P}(\boldsymbol\theta | \mathcal{S})$ and $q(\boldsymbol\theta)=\mathbb{P}(\boldsymbol\theta | \mathcal{S}^\prime)$.
The KL divergence between $p$ and $q$ is
\begin{equation}
\begin{aligned}
    \label{eq:D_kl}
    D_{KL}(p||q) &= \int_{\boldsymbol\theta} p(\boldsymbol\theta)\ln\frac{p(\boldsymbol\theta)}{q(\boldsymbol\theta)}\mathrm{d}\boldsymbol\theta 
    = \int_{\boldsymbol\theta} \frac{1}{K_{\mathcal{S}}} \exp{(-\sum_{i=1}^{\abs{\mathcal{S}}}l(\boldsymbol\theta, \mathbf{s}_1))}\ln\frac{p(\boldsymbol\theta)}{q(\boldsymbol\theta)} \mathrm{d}\boldsymbol\theta,
\end{aligned}
\end{equation}
where
\begin{equation}
\label{eq:ln_p_q}
\begin{aligned}
    \ln\frac{p(\boldsymbol\theta)}{q(\boldsymbol\theta)} &= 
\sum_{i=1}^{\abs{\mathcal{S}^\prime}} l(\boldsymbol\theta, \mathbf{s}_i^\prime) - \sum_{i=1}^{\abs{\mathcal{S}}} l(\boldsymbol\theta, \mathbf{s}_i) + K_{\mathcal{S}^\prime} - K_{\mathcal{S}} \\
&=\sum_{i=1}^{\abs{\mathcal{S}}} (l(\boldsymbol\theta, \mathbf{s}_i^\prime) -l(\boldsymbol\theta, \mathbf{s}_i)) + K_{\mathcal{S}^\prime} - K_{\mathcal{S}} \\
&\leq L \sum_{i=1}^{\abs{\mathcal{S}}}\norm{\mathbf{s}_i^\prime - \mathbf{s}_i}_2 + \abs{K_{\mathcal{S}^\prime} - K_{\mathcal{S}}}.
\end{aligned}
\end{equation}

According to the assumption \ref{assump:model_param_dist}, $K_\mathcal{S}$ (similar for $K_{\mathcal{S}^\prime}$) is:
\begin{equation}
    K_{\mathcal{S}}\coloneqq \int_{\boldsymbol\theta} \exp{(-\sum_{i=1}^{\abs{\mathcal{S}}}l(\boldsymbol\theta, \mathbf{s}_i))} \mathrm{d}\boldsymbol\theta.
\end{equation}

Since $\mathbf{x}_{\abs{\mathcal{T}}}$ is not used for real data initialization, according to the Proposition \ref{proposition:minimizer}, if $\mathcal{S}$ and $\mathcal{S}^\prime$ share the same initialization type and initialized values, we have for each $i$
\begin{equation}
\begin{aligned}
    \norm{\mathbf{s}_i^\prime - \mathbf{s}_i}_2 &= \norm{\frac{1}{\abs{\mathcal{T}}-1}\sum_{j=1}^{\abs{\mathcal{T}}-1}\mathbf{x}_j - \frac{1}{\abs{\mathcal{T}}}\sum_{j=1}^{\abs{\mathcal{T}}}\mathbf{x}_j}_2 \\
    &= \frac{1}{\abs{\mathcal{T}}} \norm{\frac{1}{\abs{\mathcal{T}}-1}\sum_{j=1}^{\abs{\mathcal{T}}-1}\mathbf{x}_j - \mathbf{x}_{\abs{\mathcal{T}}}}_2 \\
    &\leq \frac{2B}{\abs{\mathcal{T}}}.
\end{aligned}
\end{equation}

Thus, we have 
\begin{equation}
    L \sum_{i=1}^{\abs{\mathcal{S}}}\norm{\mathbf{s}_i^\prime - \mathbf{s}_i}_2 \leq \frac{2LB\abs{\mathcal{S}}}{\abs{\mathcal{T}}}.
\end{equation}

The second term on the right side of \eqref{eq:ln_p_q} can be processed similarly:
\begin{equation}
\begin{aligned}
    &\abs{K_{\mathcal{S}^\prime} - K_{\mathcal{S}}} = \abs{\int_{\boldsymbol\theta} \exp{(-\sum_{i=1}^{\abs{\mathcal{S}^\prime}}l(\boldsymbol\theta, \mathbf{s}_i^\prime))} - \exp{(-\sum_{i=1}^{\abs{\mathcal{S}}}l(\boldsymbol\theta, \mathbf{s}_i))} \mathrm{d}\boldsymbol\theta} \\
    &= \abs{\int_{\boldsymbol\theta} [\exp(\sum_{i=1}^{\abs{\mathcal{S}}}(l(\boldsymbol\theta,\mathbf{s}_i)-l(\boldsymbol\theta,\mathbf{s}_i^\prime)))-1] \exp{(-\sum_{i=1}^{\abs{\mathcal{S}}}l(\boldsymbol\theta, \mathbf{s}_i))} \mathrm{d}\boldsymbol\theta}.
\end{aligned}
\end{equation}

From previous analysis, we know that
\begin{equation}
    \sum_{i=1}^{\abs{\mathcal{S}}}(l(\boldsymbol\theta,\mathbf{s}_i)-l(\boldsymbol\theta,\mathbf{s}_i^\prime)) \leq \frac{2LB\abs{\mathcal{S}}}{\abs{\mathcal{T}}}.
\end{equation}

Since $\exp(x)-1 = O(x)$ in the neighborhood of $0$, we have 
\begin{equation}
    \abs{K_{\mathcal{S}^\prime} - K_{\mathcal{S}}} = O(\frac{2LB\abs{\mathcal{S}}K_{\mathcal{S}}}{\abs{\mathcal{T}}}) = O(\frac{\abs{\mathcal{S}}}{\abs{\mathcal{T}}}).
\end{equation}

Note that $K_\mathcal{S}$ should decrease as $\abs{\mathcal{S}}$ increases because an additional synthetic sample $\mathbf{s}$ introduces a factor $\exp(-l(\btheta, \mathbf{s}))\leq 1$ in the integral.
We omit it here and assume $K_\mathcal{S}$ varies little when $\abs{\mathcal{S}}$ changes.
Together with \eqref{eq:D_kl} and \eqref{eq:ln_p_q}, we obtain the privacy bound by KL divergence:
\begin{equation}
\boxed{D_{KL}(p||q) = O(\frac{\abs{\mathcal{S}}}{\abs{\mathcal{T}}}).}
\end{equation}

\section{Generalization to Non-Linear Extractor}
\label{appendix:non_linear_extractor}

We consider 2-layer network as the extractor, \ie, linear extractor with ReLU activation, and show that the (pseudo)-barycenters of $\mathcal{S}$ and $\mathcal{T}$ also coincide as claimed by Proposition \ref{proposition:DM_linear}.
We then empirically validate the conclusion by plotting the $L_2$ distance between $\mathcal{S}$ and $\mathcal{T}$ during the condensation by DM for different $r_{ipc}$ on CIFAR-10 (see Figure \ref{fig:verify_proposition_DM_linear}).

\subsection{Analysis for 2-layer Network as Extractor}
With activation function ReLU (noted as $\rho$), Equation \eqref{eq:d_DM} becomes:
\begin{equation}
   \label{eq:d_DM_relu}
   \mathbf{d}_{DM}^{ReLU} \coloneqq   \frac{1}{\abs{\mathcal{T}}}\sum\limits_{i=1}^{\abs{\mathcal{T}}} \rho(\btheta \cdot\mathbf{x}_i) - \frac{1}{\abs{\mathcal{S}}}\sum\limits_{i=1}^{\abs{\mathcal{S}}} \rho(\btheta \cdot\mathbf{s}_i).
\end{equation}
Since $\theta_{i,j}  \stackrel{iid}{\sim}\mathcal{N}(0, 1)$ for each element $\theta_{i,j}$ of $\btheta$, for an input $\mathbf{x}=[x_j]_{1\leq j \leq d}\in\mathbb{R}^d$, we have
\begin{equation}
    \mathbf{y}=\btheta\cdot\mathbf{x} = [\sum\limits_{j=1}^{d}\theta_{i,j}x_j]_{1\leq i \leq k}=[y_i]_{1\leq i\leq k}\in\mathbb{R}^k,
\end{equation}
where $y_i\stackrel{iid}{\sim}\mathcal{N}(0, \sum_{j=1}^{d}x_j^2)$.
Since $\rho(x)\coloneqq\max(0, x)$, we have $\rho(\mathbf{y}) = [\max(0, y_i)]_{1\leq i \leq k}$.
Define $Y=\max(0, X)$ where the random variable $X\sim\mathcal{N}(0, \sigma^2)$.
Then, $Y$ follows the same distribution of $B\abs{X}$, where $B\sim Bernoulli(\frac{1}{2})$ independent of $X$ and $\mathbb{E}_{X}[Y] = \mathbb{E}_{B}[B]\mathbb{E}_{X}[\abs{X}]$.
Therefore, for each $i$, $\max(0, y_i) = B_i \abs{y_i} = B_i \abs{\sum\limits_{j=1}^{d}\theta_{i,j}x_j}$, and we can obtain
\begin{equation}
    \rho(\mathbf{y}) = \rho(\btheta \mathbf{x}) = \mathbf{B} \odot \abs{\btheta \mathbf{x}}
\end{equation}
where $\odot$ is element-wise multiplication, $\mathbf{B} = [B_i]_{1\leq i\leq k}$ and $B_i\stackrel{iid}{\sim}Bernoulli(\frac{1}{2})$.
With this in mind, Equation \eqref{eq:d_DM_relu} becomes:
\begin{equation}
\begin{aligned}
    \mathbf{d}_{DM}^{ReLU} &=   \frac{1}{\abs{\mathcal{T}}}\sum\limits_{i=1}^{\abs{\mathcal{T}}} \rho(\btheta \cdot\mathbf{x}_i) - \frac{1}{\abs{\mathcal{S}}}\sum\limits_{i=1}^{\abs{\mathcal{S}}} \rho(\btheta \cdot\mathbf{s}_i) =\frac{1}{\abs{\mathcal{T}}}\sum\limits_{i=1}^{\abs{\mathcal{T}}} \mathbf{B}^x_i \odot \abs{\btheta \mathbf{x}_i} - \frac{1}{\abs{\mathcal{S}}}\sum\limits_{i=1}^{\abs{\mathcal{S}}} \mathbf{B}^s_i \odot \abs{\btheta \mathbf{s}_i},
\end{aligned}
\end{equation}
where vectors of Bernoulli random variable for each data samples $\mathbf{B}_i^\cdot$ are independent.
To simplify notation, we consider $k=1$.
The vector of Bernoulli random variable reduces to single random variable $B_i^\cdot$, and the bold symbol $\mathbf{d}$ becomes $d$.
Moreover, let $\text{sgn}(x)$ denote the sign of $x$, and we can see that $\abs{x}=\text{sgn}(x) x$ for a real number $x$.
Thus, with $k=1$, we can reduce $d_{DM}^{ReLU}$ to the similar form of \eqref{eq:d_DM}:
\begin{equation}
\begin{aligned}
     d_{DM}^{ReLU} &= \frac{1}{\abs{\mathcal{T}}}\sum\limits_{i=1}^{\abs{\mathcal{T}}}  B^x_i\text{sgn}(\btheta \mathbf{x}_i)\btheta \mathbf{x}_i - \frac{1}{\abs{\mathcal{S}}}\sum\limits_{i=1}^{\abs{\mathcal{S}}}  B^s_i\text{sgn}(\btheta \mathbf{s}_i)\btheta \mathbf{s}_i \\
     &=\btheta(\frac{1}{\abs{\mathcal{T}}}\sum\limits_{i=1}^{\abs{\mathcal{T}}}  B^x_i\text{sgn}(\btheta \mathbf{x}_i) \mathbf{x}_i - \frac{1}{\abs{\mathcal{S}}}\sum\limits_{i=1}^{\abs{\mathcal{S}}}  B^s_i\text{sgn}(\btheta \mathbf{s}_i)\mathbf{s}_i).
\end{aligned}
\end{equation}
Recall that for each $j$, $\partial L_{DM} /\partial \mathbf{s}_j = \mathbb{E}_{\btheta}[(d_{DM}^{ReLU})^2 /\partial \mathbf{s}_j] = \mathbb{E}_{\btheta}[2 (\partial d_{DM}^{ReLU}/ \partial \mathbf{s}_j)d_{DM}^{ReLU}]$, and we have
\begin{equation}
    \frac{\partial d_{DM}^{ReLU}}{\partial \mathbf{s}_j} = -\frac{1}{\abs{\mathcal{S}}}B_j^s\text{sgn}(\btheta\mathbf{s}_j)\btheta^\top.
\end{equation}
Thus, the gradient of $L_{DM}$ on $\mathbf{s}_j$ becomes:
\begin{equation}
\begin{aligned}
    \frac{L_{DM}}{\mathbf{s}_j} &=\mathbb{E}_{B,\btheta}[ -\frac{2}{\abs{\mathcal{S}}} \btheta^\top\btheta (\frac{1}{\abs{\mathcal{T}}}\sum\limits_{i=1}^{\abs{\mathcal{T}}} B_j^s\text{sgn}(\btheta\mathbf{s}_j) B^x_i\text{sgn}(\btheta \mathbf{x}_i) \mathbf{x}_i - \frac{1}{\abs{\mathcal{S}}}\sum\limits_{i=1}^{\abs{\mathcal{S}}} B_j^s\text{sgn}(\btheta\mathbf{s}_j) B^s_i\text{sgn}(\btheta \mathbf{s}_i)\mathbf{s}_i)] \\
    &=-\frac{2}{\abs{\mathcal{S}}}  (\frac{1}{\abs{\mathcal{T}}}\sum\limits_{i=1}^{\abs{\mathcal{T}}} \mathbb{E}_B[B_j^s B^x_i]\mathbb{E}_{\btheta}[\text{sgn}(\btheta\mathbf{s}_j) \text{sgn}(\btheta \mathbf{x}_i)\btheta^\top\btheta ] \mathbf{x}_i - \frac{1}{\abs{\mathcal{S}}}\sum\limits_{i=1}^{\abs{\mathcal{S}}} \mathbb{E}_B[B_j^s B^s_i]\mathbb{E}_{\btheta}[\text{sgn}(\btheta\mathbf{s}_j) \text{sgn}(\btheta \mathbf{s}_i)\btheta^\top\btheta ]\mathbf{s}_i).
\end{aligned}
\end{equation}
Let $M(\mathbf{x}, \mathbf{y})$ denote $ \mathbb{E}_{\btheta}[\text{sgn}(\btheta\mathbf{x}) \text{sgn}(\btheta \mathbf{y})\btheta^\top\btheta]\in \mathbb{R}^{d\times d}$, then the above equation becomes:
\begin{equation}
\boxed{
    \frac{L_{DM}}{\mathbf{s}_j} = -\frac{2}{\abs{\mathcal{S}}}  (\frac{1}{\abs{\mathcal{T}}}\sum\limits_{i=1}^{\abs{\mathcal{T}}} \frac{1}{4} M(\mathbf{s}_j, \mathbf{x}_i) \mathbf{x}_i - \frac{1}{\abs{\mathcal{S}}}\sum\limits_{i=1, i\neq j}^{\abs{\mathcal{S}}} \frac{1}{4} M(\mathbf{s}_j, \mathbf{s}_i)\mathbf{s}_i - \frac{1}{2\abs{\mathcal{S}}}\mathbf{s}_j),}
\end{equation}
because $M(\mathbf{x}, \mathbf{y})=\mathbf{I}_d$ if $\mathbf{x} = \mathbf{y}$.
Note that if $\mathbf{x} = -\mathbf{y}$, then $M(\mathbf{x}, \mathbf{y})=-\mathbf{I}_d$.
In fact, we can prove that
\begin{equation}
    \mathbf{x}^\top M(\mathbf{x}, \mathbf{y}) \mathbf{y} = \mathbb{E}_{\btheta}[(\text{sgn}(\btheta\mathbf{x})\btheta\mathbf{x})^\top (\text{sgn}(\btheta \mathbf{y})\btheta\mathbf{y})] = \frac{\norm{\mathbf{x}}_2 \norm{\mathbf{y}}_2}{\pi}[(\pi-2\phi)\cos(\phi) + 2\sin(\phi)],
\end{equation}
which can be seen as a matrix depending on the angle $\phi$ between $\mathbf{x}$ and $\mathbf{y}$.
Even though each original data $\mathbf{x}_i$ is varied by $M(\mathbf{s}_j, \mathbf{x}_i)$, their average can still be seen as a pseudo-barycenter, and the above equation signifies that each $\mathbf{s}_j$ is updated towards minimizing the distance between the pseudo-barycenters of $\mathcal{T}$ and $\mathcal{S}$, which verifies the first property of Proposition \ref{proposition:DM_linear} on non-linear extractor.
This further validates the privacy property of DM which is based on the connection between $\mathcal{S}$ and $\mathcal{T}$.

Next, we empirically verify the Proposition \ref{proposition:DM_linear} with tests on CIFAR-10.

\subsection{Empirical verification of Proposition \ref{proposition:DM_linear} for non-linear extractor}
\begin{figure}
    \centering
    \resizebox{0.9\linewidth}{!}{\includegraphics{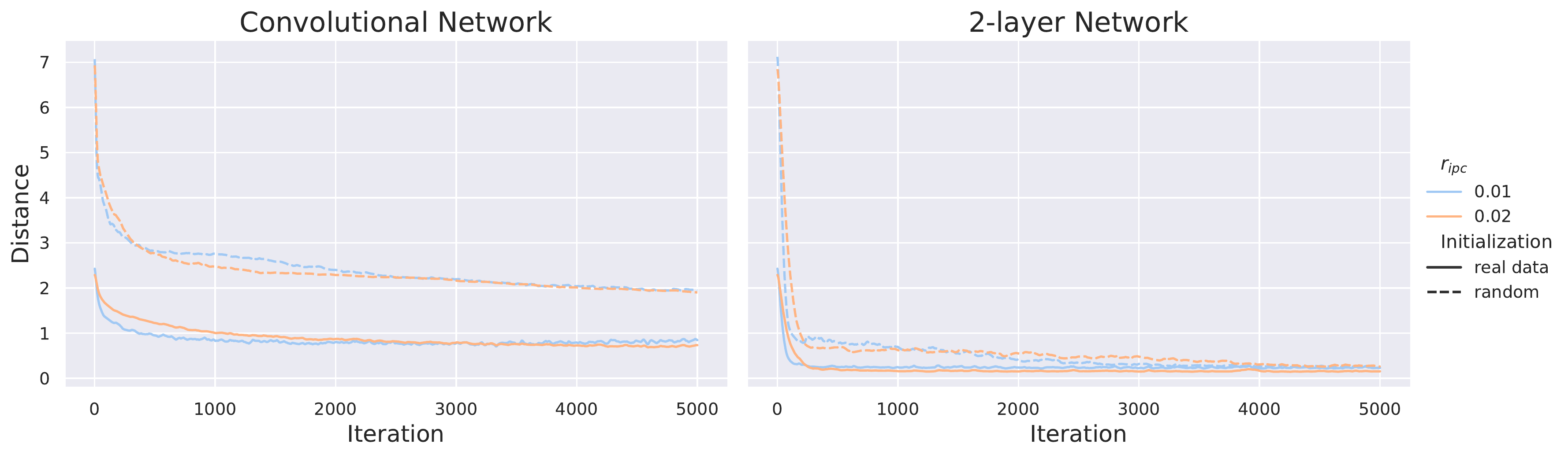}}
    \caption{Distance ($\norm{\cdot}_2$) between barycenters of $\mathcal{S}$ and $\mathcal{T}$ deceases with the iteration round, which verifies first property of Proposition \ref{proposition:DM_linear}.
    The solid and dashed lines represent real data and random initialization, respectively.
    The blue and orange lines represent the cases where $r_{ipc}$ equals to $0.01$ and $0.02$, respectively.}
    \label{fig:verify_proposition_DM_linear}
\end{figure}
Figure \ref{fig:verify_proposition_DM_linear} shows the distance $\norm{\frac{1}{\abs{\mathcal{S}}}\sum_{i=1}^{\abs{\mathcal{S}}}\mathbf{s}_i - \frac{1}{\abs{\mathcal{T}}}\sum_{i=1}^{\abs{\mathcal{T}}}\mathbf{x}_i}_2$ for each DM iteration on CIFAR-10.
We can observe that the barycenter distance decreases with the iteration round, and achieves to the minimum.
Note that the right subfigure of Figure \ref{fig:verify_proposition_DM_linear} shows that the barycenters of $\mathcal{T}$ and $\mathcal{S}$ synthesized on 2-layer network (\ie, linear model activated by ReLU) have distance around $0$, validating the theoretical analysis above.
As for convolutional network (ConvNet), the distance decreases slower than 2-layer network.
We suspect that the convolutional layers will lead the optimization to a local minimum.
Figure \ref{fig:verify_proposition_DM_linear} also validates the impact of $r_{ipc}$ and initialization to the distance of barycenters of $\mathcal{S}$ and $\mathcal{T}$: 1) when the iteration round is around $0$, the distance of $100$ image per class is smaller than that of $50$ image per class, 2) the real initialization has much lower distance than of random initialization at the beginning of DM optimization.

\section{Additional Experimental Details and Results}
\label{appendix:additional_exp}
All experiments are conducted with Pytorch 1.10 on a Ubuntu 20.04 server.
\subsection{Details of Hyperparameters and Settings. }
\label{appendix:settings}

\textbf{DC Settings.} We reproduced DM  
\cite{DC_distribution_matching} and adopt large learning rates to accelerate the condensation (\ie, $10, 50, 100$ as learning rate for $r_{ipc}=0.002, 0.01, 0.02$, respectively).
For DSA \cite{DSA}, we adopt the default setting \footnote{\url{https://github.com/VICO-UoE/DatasetCondensation}} for all datasets.
For KIP, we reproduced in Pytorch according to the official code of \cite{KIP_iclr21}, and set learning rate $0.04$ and $0.1$ for $r_{ipc}=0.002$ and $0.01$, respectively.
Note that we omit $r_{ipc}=0.02$ for KIP and DSA due to the low efficiency.
We also apply differentiable siamese augmentations \cite{DSA} for both DM and KIP.


\subsection{Loss distribution of data used for DC initialization and test data on $f_\mathcal{S}$}
\label{appendix:loss_dist_dc_with_real_init}
\begin{figure*}
    \centering
    \resizebox{0.6\linewidth}{!}{
    \includegraphics{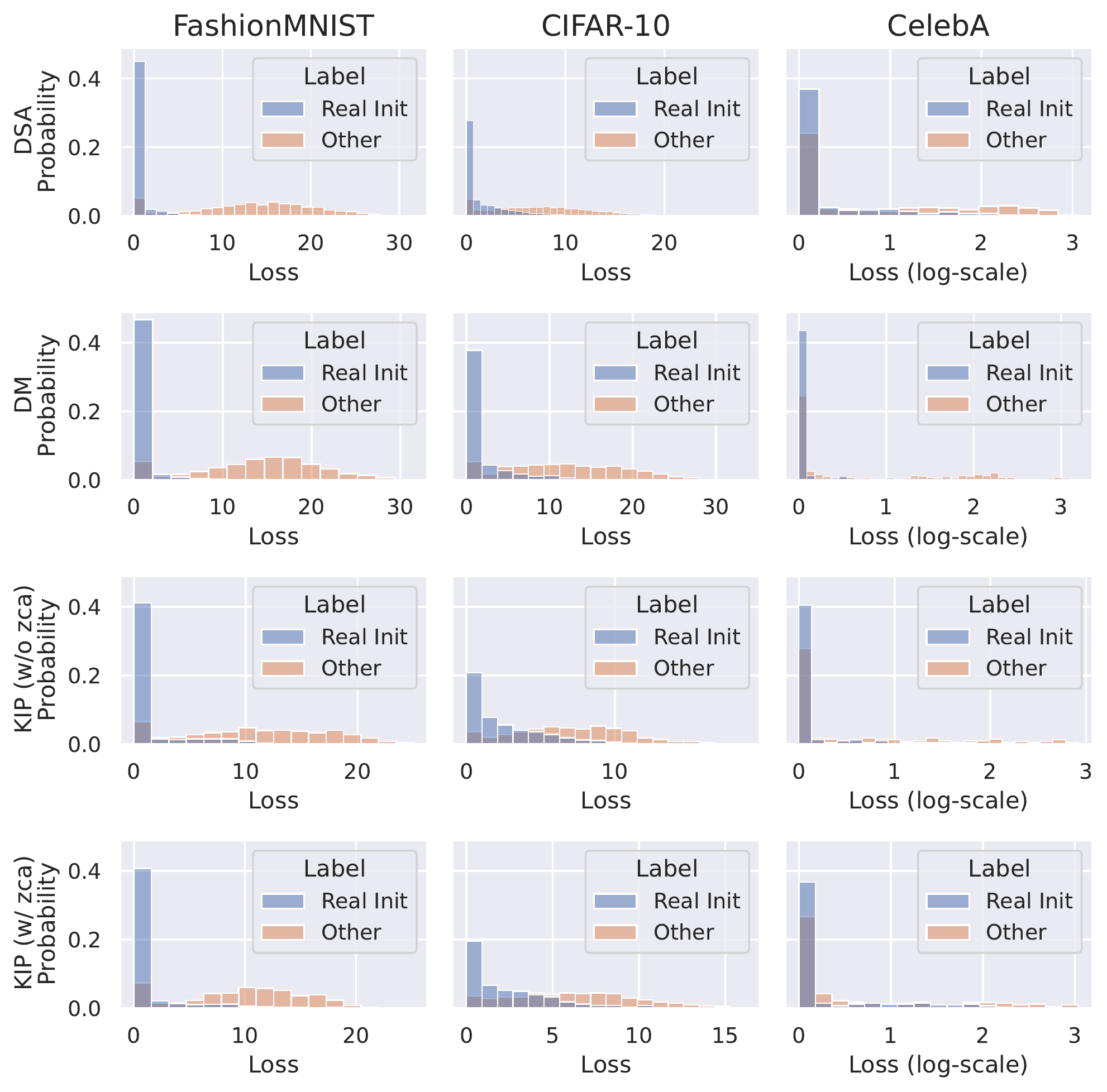}}
  \caption{Loss distribution of data used for DC initialization (Real Init) are smaller than data not used for initialization (Other). 
    }
    \label{fig:real_init_hist}
\end{figure*}
In Figure \ref{fig:real_init_hist}, we show the distribution of $f_\mathcal{S}$ losses evaluated on data used for DC initialization (Real Init) and data not used for initialization (Other).
We can observe that the losses of data used for initialization are smaller than other data, showing that the membership of data used for DC initialization are easier to be inferred.
The distribution difference also explains the high advantage scores in Table \ref{tab:real_init_advantage}.

\subsection{Visualization of DC-synthesized data distribution}
\label{subsec:visual_distribution}
\begin{figure*}
    \centering
    \resizebox{0.65\linewidth}{!}{\includegraphics{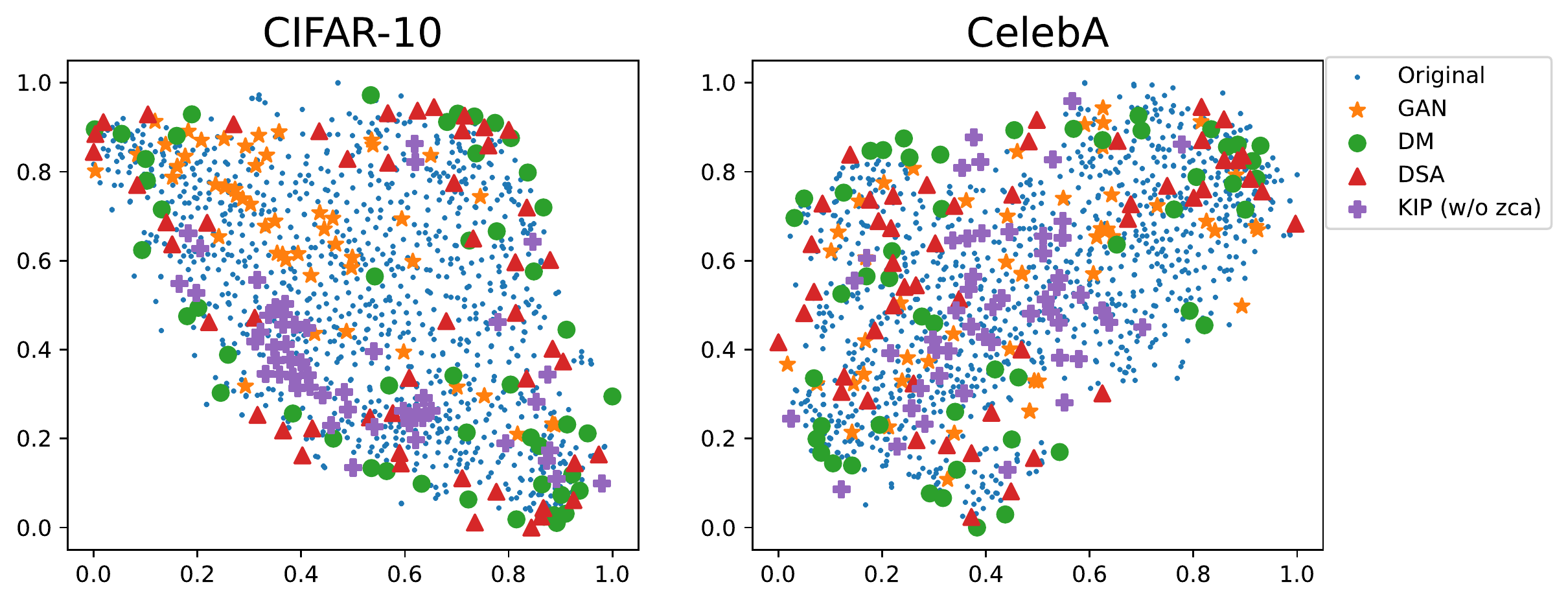}}
    \caption{Distribution visualization of CIFAR-10 (left) and CelebA (right) synthesized by GAN, DSA, DM and KIP (without ZCA preprocessing).}
    \label{fig:visual_distribution}
\end{figure*}
Figure \ref{fig:visual_distribution} shows the t-SNE visualization of CIFAR-10 and CelebA data synthesized by GAN and DC methods (DSA, DM and KIP without ZCA preprocessing).
We clip the DC-synthesized into $0$ and $1$ for fair comparison with GAN-synthesized data.
Note that the generated data distributions of DM and DSA are more similar than KIP and GAN, explaining why DM-synthesized data and DSA-synthesized data enable models to achieve higher accuracy under same $r_{ipc}$.

\subsection{MIA against cGANs}
\label{appendix:mia_againt_gan}
Our threat model assumes that the adversary has white-box access to the synthetic dataset.
We apply the MIA against GANs proposed by \citeauthor{DBLP:conf/ccs/ChenYZF20} (called GAN-leak).
The main intuition is that member data are easier to be reconstructed by GAN generators $\mathcal{G}$, so the MIA is based on the (calibrated) reconstructed loss $L_{cal}$:
\begin{equation}
    M(\mathbf{x}) = \mathbbm{1}(L_{cal}(\mathbf{x}, \mathcal{G}(\mathbf{z}))\leq \tau).
\end{equation}
The adversary optimizes $L_{cal}$ by varying $\mathbf{z}$ to estimate whether $\mathbf{x}$ belongs to the training dataset.
According to the adversary' knowledge, the attack can be divided into black-box attack, partial black-box attack and white-box attack.
We conducted the white-box attack for scenarios where the adversary has access to the generators.
The results on CelebA are in Table \ref{tab:ganleak_results}, indicating that vanilla GAN can be used to infer the membership of training data.
\citeauthor{DBLP:conf/ccs/ChenYZF20} also validated that partial black-box attack can achieve similar attack performance as white-box, because the adversary has access to $\mathbf{z}$ and can leverage non-differentiable optimization, \eg, the Powell’s Conjugate Direction Method \cite{DBLP:journals/cj/Powell64}), to approximately minimize $L_{cal}$.

\begin{table}[htbp]
\centering
\caption{Results of GAN-leak attack against cGANs averaged over $10$ shadow models.}
\label{tab:ganleak_results}
\begin{tabular}{ccc}
\toprule
\textbf{Dataset} & ROC AUC & Advantage (\%) \\ \midrule
CelebA & $56.06\pm 2.03$ & $22.98\pm 4.27$  \\
 \bottomrule
\end{tabular}
\end{table}

\subsection{Comparison of accuracy for models trained on synthetic dataset for $r_{ipc}=0.002$}
\begin{figure*}
    \centering
    \resizebox{1\linewidth}{!}{\includegraphics{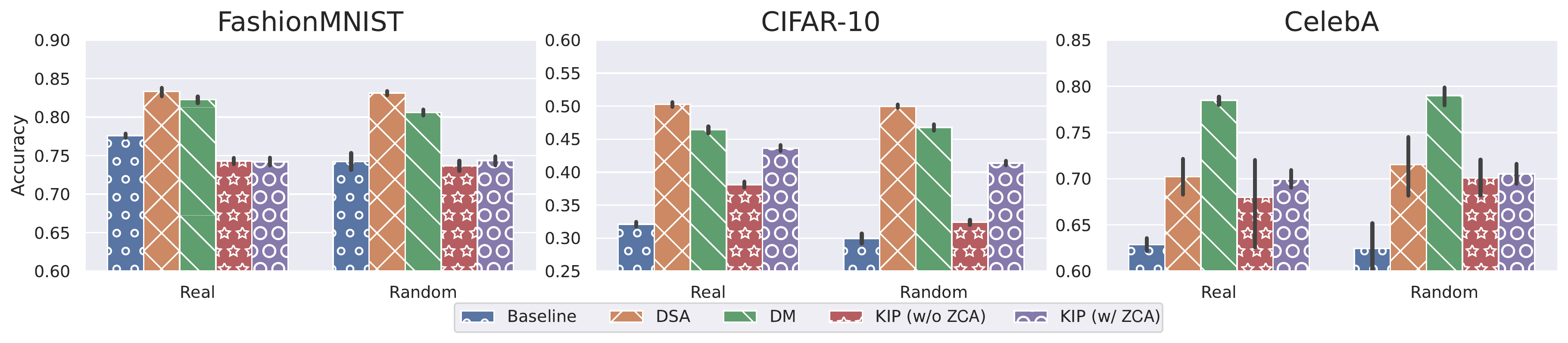}}
    \caption{Accuracy of models trained on data synthesized by different DC methods and on data generated by baselines for $r_{ipc} = 0.002$.}
    \label{fig:accuracy_barplot0.002}
\end{figure*}

Figure \ref{fig:accuracy_barplot0.002} presents the accuracy comparison results of models trained on data synthesized by DC and baseline methods for $r_{ipc}=0.002$.
We can see that KIP significantly outperforms baselines and achieves similar performance with DSA and DM on CIFAR-10.
Moreover, we can observe that the ZCA preprocessing is effective for improving the utility of KIP-synthesized dataset.

\end{document}